\documentclass[10pt,aps,pra,epsfigure,notitlepage,twocolumn,superscriptaddress]{revtex4-1}

\usepackage{bm} 
\usepackage{graphicx}
\usepackage{amsmath}    
\usepackage{latexsym}
\usepackage{amsfonts}   
\usepackage{amssymb}
\usepackage{comment}
\usepackage{array}      

\usepackage{txfonts}
\usepackage{xcolor}
\usepackage{float}
\usepackage{nameref}
\usepackage{placeins}
\usepackage[colorlinks=true,linkcolor=blue,urlcolor=blue,citecolor=blue,pdfusetitle]{hyperref}

\newcommand {\fabs}[1] {\left| #1 \right|}

\newcommand{\ket}[1]{\ensuremath{|#1\rangle}}
\newcommand{\bra}[1]{\langle#1|}

\newcommand{\ketbra}[2]{|#1\rangle\langle#2|}

\usepackage{ulem}

\newcommand{\steve}[1]{\textcolor{black}{#1}}

\usepackage{float}

\begin{document}
\title{Ergotropy transport in a one dimensional spin chain}
\author{Dara Murphy}
\affiliation{School of Physics, University College Dublin, Belfield, Dublin 4, Ireland}
\affiliation{Centre for Quantum Engineering, Science, and Technology, University College Dublin, Dublin 4, Ireland}

\author{Anthony Kiely}
\affiliation{School of Physics, University College Dublin, Belfield, Dublin 4, Ireland}
\affiliation{Centre for Quantum Engineering, Science, and Technology, University College Dublin, Dublin 4, Ireland}
\affiliation{School of Physics, University College Cork, College Road, Cork, Ireland}

\author{Irene D'Amico}
\address{School of Physics, Engineering and Technology, University of York, York, United Kingdom}
\address{York Centre for Quantum Technologies, University of York, York, United Kingdom}

\author{Steve Campbell}
\affiliation{School of Physics, University College Dublin, Belfield, Dublin 4, Ireland}
\affiliation{Centre for Quantum Engineering, Science, and Technology, University College Dublin, Dublin 4, Ireland}

\begin{abstract}
We examine the transport of useful energy, i.e. extractable work as quantified by the ergotropy, along a spin chain with tuneable exchange couplings between the sites. We focus on, and interpolate between, the two physically relevant limits of uniform interaction strengths and engineered couplings which achieve  perfect state transfer (PST). By modelling the individual constituents as quantum batteries, we consider how the manner in which the extractable work appears in the initial state of the first site impacts the chain's ability to transport ergotropy to the final site. For non-PST couplings, we establish that there is a clear quantum advantage when the ergotropy is initially endowed in quantum coherences and demonstrate that this ergotropy is more efficiently transferred. For extractable work encoded in a population inverted state, we show that this considerably limits the length of chain over which any ergotropy can be faithfully transported. For PST couplings, we consider the robustness to disorder and again demonstrate a quantum advantage for coherently endowed ergotropy. Finally, we examine the work probability distribution associated with quenching on the interactions which provides insight into the work cost in switching on the couplings. We show that PST couplings lead to smaller fluctuations in this work cost, indicating that they are more stable.
\end{abstract}
\maketitle

\section{Introduction}
The formulation of thermodynamic laws for nanoscale systems has progressed substantially in the last decades~\cite{Binder2019book, Campbell2025roadmap, Gherardini2024}. From a foundational standpoint, examining how concepts and intuitions must be augmented when the peculiarities of quantum mechanics are considered has provided valuable insights, for example sharpening Landauer's principle connecting information and energy~\cite{Reeb2014} and quantum generalisations of the second law~\cite{Brand_o2015,Strasberg2021}, among others. Recently, however, there has been a growing need to critically assess the thermodynamics and energetics of emerging quantum technologies in order to ensure their scalability~\cite{Campbell2025roadmap, Auffeves2022}. 

There is a subtle distinction between energy and useful energy, i.e. work. For a system in a given state, $\rho$, with Hamiltonian, ${\mathcal{H}}$, we can define the amount of useful energy the system possesses via the ergotropy
\begin{eqnarray}
\mathcal{E}(\rho)& = &\operatorname{Tr} \left[ \rho {\mathcal{H}} \right] - \underset{U}{\text{min}} \operatorname{Tr} \left[  U_{\tau} \rho U_{\tau}^{\dagger} {\mathcal{H}} \right] \nonumber \\ &=& \operatorname{Tr} \left[ {\rho \mathcal{H}} \right] - \operatorname{Tr} \left[ {\mathcal{H}} \rho_p \right],
\label{eq:Ergotropy}
\end{eqnarray}
which is the maximal amount of extractable work from the system after undergoing a cyclic unitary process~\cite{Allahverdyan2004, Alicki2013PRE}. The minimisation is taken over all unitaries, $U_{\tau}$, such that it leaves the state of the system in its lowest available energy configuration, i.e. a passive state $\rho_p$. A battery is therefore any system that stores a non-zero amount of extractable work. Viewing a quantum mechanical system as a battery allows us to consider the impact that genuine quantum features, in particular coherence, can have on our understanding of how the laws of thermodynamics must be adapted at the nano-scale~\cite{Campaioli2024Colloquium}. Several works have explored the impact of coherence and correlations on ergotropy~\cite{binder2015quantacell, Hovhannisyan2013PRL, giorgi2015correlation, QCohErg, Cakmak2020PRE, Kamin2020PRE, NthRoot,mondal2025,Tirone2021}, designed schemes for stable and efficient charging and discharging protocols~\cite{Santos2019PRE, Quach2020PRApp, moraes2022charging, Barra20219PRL, evangelakos2025rapid, satriani2024daemonic, deMoraes2024quantum, Rinaldi2025PRA,Tirone2023}, and explored how different microscopic models behave when modelled as a quantum battery~\cite{Le2018Spin, Hoang2024, razzoli2025cyclic}. Quantum batteries thus provide a useful framework to explore both fundamental aspects of the energetics of quantum systems while also being of potential practical relevance for newly emerging quantum devices~\cite{Li2025, quach2022superabsorption, tibben2025PRXEnergy}. Such advances notwithstanding, recent focus has shifted to exploring the transfer of useful work. This becomes relevant when one considers that the means by which a quantum battery reaches an active state inevitably involves some interaction with a charging system~\cite{Andolina2018PRB}. It has also been shown that the presence of genuine quantum features can ensure for more efficient transfer of work between quantum battery and a consumption unit~\cite{CorrLossErg}.

In this work we complement this line of inquiry by considering a closely related process: the shuttling of extractable work along a spin chain. The use of spin-systems as quantum communication channels is well established~\cite{QcommUnmod, QcommReview, Kandel2021, Qiao2021}, with protocols designed to achieve, for example, high-fidelity state transfer and correlation transfer, see e.g.~\cite{kay2010perfect, PSTQCA, PSTarbQCA, bezaz2025quasi, alsulami2024scalable, nikolopoulos2014quantum, nikolopoulos2004electron, Estarellas2017PRA, DiFranco2008PRL, Campbell2011PRA, Apollaro2012PRA, banchi2011long, Vinet2012PRA, wilkinson2017rapid}. While the information theoretic aspects have been extensively studied, the examination of the thermodynamics of these communication channels is still in its infancy~\cite{LewisPRB2023}. The lack of such an analysis may, in part, be due to the fact that for the often considered energy-preserving models that will be the focus of this work, certain properties are communicated identically, e.g. spin or energy. Therefore, one might naively assume that other thermodynamic quantities, such as the ergotropy, would behave similarly. In this work we demonstrate that extractable work is distinct and the ability for a spin chain to effectively transfer ergotropy is delicately dependent on the manner in which it is initially endowed. In particular, we demonstrate that a quantum advantage emerges in the presence of initial coherences. Furthermore, we also assess the thermodynamics associated with switching on the interactions across the chain showing that, while the average work cost can be zero, the nature of the coupling impacts the fluctuations of the work done. In particular, while intrinsically more difficult to engineer, couplings that achieve PST are shown to be more stable at the level of the full work distribution.

\section{Model}
\label{model}
We consider a chain of spin-1/2 particles, each with a local Hamiltonian $H\!=\! - B \sigma_{z}^{i}$ where $B$ sets the onsite energy splitting. We will assume that the spin at the first site is in an active state, i.e. it possesses some extractable work, which we aim to transfer along the chain. To facilitate this, at time $t\!=\!0$ exchange interactions between the spins are suddenly switched on and the total system is then described by a 1-dimensional spin-1/2 chain of length $N$ with open boundary conditions. The system's evolution is governed by the Hamiltonian,
\begin{equation}
\mathcal{H}= \sum_{j=1}^{N-1} \mathcal{J}_\alpha \left( \sigma_{x}^{j} \sigma_{x}^{j+1} + \sigma_{y}^{j} \sigma_{y}^{j+1} \right) - B \sum_{j=1}^N \sigma_z^j,
\label{eq:Hxx}
\end{equation}
where $\mathcal{J}_\alpha\!=\!\frac{1}{2}[ (\alpha-1)J+\alpha J_{j,j+1}]$. Due to the excitation preserving nature of the interactions we can restrict to the zero- and single-excitation subspace only. We will therefore find it convenient to work using the basis states $\ket{\underline{\mathbf{1}}}_n \!=\! \bigotimes_{j=1}^{n-1} \ket{0}_j \otimes \ket{1}_n \bigotimes_{l=n+1}^N \ket{0}_l$ which corresponds to the $n^{\rm th}$ site of the chain in its local excited state, and $\ket{\underline{\mathbf{0}}}\!=\! \bigotimes_{j=1}^N \ket{0}_j$ corresponding to all spins in the chain to be in their local ground state. The parameter $\alpha$ allows us to interpolate between uniform couplings ($\alpha\!=\!0$) and the couplings required to achieve perfect state transfer (PST)~\cite{PSTarbQCA,PSTQCA} ($\alpha\!=\!1)$. The PST coupling strengths are given by,
\begin{equation}
\label{eq:PSTcoupling}
    J_{j,j+1} =  \frac{2J}{N} \sqrt{j(N - j)} G_N
\end{equation}
where
\begin{eqnarray}
    G_N=\begin{cases}
        \displaystyle 1 , & \text{if } N \text{ is even}, \\[8pt]
        \displaystyle \frac{1}{\sqrt{1 - \frac{1}{N^2}}}, & \text{if } N \text{ is odd},
    \end{cases}.
\end{eqnarray}

For the limiting cases, $\alpha\!=\!0$ or 1, we can find the spectrum of $\mathcal{H}\!=\!\sum_{k=0}^N E_k \ketbra{E_k}{E_k}$. For $\alpha\!=\!1$, the eigenenergies are,
\begin{equation}
E_k = 
\begin{cases}
    -N B, &  k = 0, \\
    -\dfrac{2J}{N}[N - (2k - 1)] G_N - (N - 2) B, & k > 0,
\end{cases}
\label{eq:EigvalsCA}
\end{equation}
with eigenstates,
\begin{equation}
\ket{E_k} = 
\begin{cases}
    \ket{\underline{\mathbf{0}}}, &  k = 0, \\
    \displaystyle\sum_{n=1}^{N} \sqrt{w(n)}\, K_k\Bigl(n; \tfrac{1}{2}, N\Bigr)  \ket{\underline{\mathbf{1}}}_n, &  k > 0,
\end{cases}
\label{eq:EigvecsCA}
\end{equation}
where $w(n)$ is the weight function for the Krawtchouk polynomial $K_k\Bigl(n; \tfrac{1}{2}, N\Bigr)$.

For $\alpha\!=\!0$, corresponding to a chain with uniform coupling strengths, the energy eigenvalues take the form~\cite{PSTarbQCA},
\begin{equation}
    E_k = \begin{cases}
 -NB, & k=0, \\
-2J\cos\left({\frac{\pi k}{N+1}}\right)-(N-2)B, & k>0,
\end{cases}
\label{eq:EigvalsXX}
\end{equation}
with eigenstates,
\begin{equation}
\ket{E_k} = 
\begin{cases}
    \ket{\underline{\mathbf{0}}}, &  k = 0, \\
    \sqrt{\dfrac{2}{N+1}} \displaystyle\sum_{n=1}^N \sin\left( \dfrac{\pi k n}{N+1} \right) \ket{\underline{\mathbf{1}}}_n\, , &  k > 0.
\end{cases}
\label{eq:EigvecsXX}
\end{equation}
In general, uniform couplings do not allow for PST (with the notable exception being extremely small chain sizes of $N\leq3$). Nevertheless, uniform couplings do allow for {\it some} transport. The parameter $J$ sets the overall energy scale for our setup. For $\alpha\in [0,1]$, the width of the energy spectrum of Eq.~\eqref{eq:Hxx} is bounded in the interval $[-2J,2J]$ centred on, $-(N-2)B$.  In Appendices~\ref{CA} and \ref{XX} we provide details of the derivations for both limits  and refer the interested reader to Refs.~\cite{Albanese2004,PSTarbQCA,inhomspinchain,Vinet2012PRA} for a more complete analysis.

Ergotropy can be stored in a system in two ways: either through a population inversion and/or in coherences in the energy eigenbasis. When extractable work is present, encoded in either form, the system is said to be in an active state~\cite{Allahverdyan2004}. Our interest will be on demonstrating that initially storing ergotropy in quantum coherences is highly advantageous with respect to the ability to shuttle it along the chain when the couplings are not identically those which achieve PST. We therefore consider the following initial states for the chain,
\begin{eqnarray}
\ket{\psi(0)} &=& \left(\cos\tfrac{\theta}{2} \ket{0}_1 + e^{i\phi}\sin\tfrac{\theta}{2} \ket{1}_1 \right) \bigotimes_{i=2}^N \ket{0}_i \label{pureinitial} \, , \\
\rho(0) &=& \big( q \ket{1} \bra{1}_1 + (1-q)\ket{0} \bra{0}_1 \big) \bigotimes_{i=2}^N \ket{0} \bra{0}_i. \label{mixedinitial}
\end{eqnarray}
In Eq.~\eqref{pureinitial} we see the first site is initially pure and, therefore, the initial ergotropy to be transported is endowed in both the populations and coherence. In contrast, the first site in Eq.~\eqref{mixedinitial} is a mixed state in its local energy eigenbasis. Ergotropy can be encoded in this state also, however it requires a population inversion, i.e. $q\!>\!0.5$. To ensure a fair comparison, we will consider states which contain the same amount of initial ergotropy which, by directly computing Eq.~\eqref{eq:Ergotropy} for the first site of the chain, corresponds to the condition $q\!=\!\left[1+\sin^2\left(\frac{\theta}{2}\right)\right]/2$. Finally, we remark that although $\phi$ in Eq.~\eqref{pureinitial}, plays a significant role in a system's dynamics, e.g fidelity, it does not play a role on the level of the energetics~\cite{Bayat2011}. Therefore, without loss of generality we set $\phi=0$.

\section{Dynamics of Extractable Work}
\label{sec: Figures of Merit}
\subsection{Evolved state and transition probabilities}
The choice of initial states in Eqs.~\eqref{pureinitial} and \eqref{mixedinitial} and excitation preserving form of Eq.~\eqref{eq:Hxx} means that we can restrict our analysis to a zero- and single-excitation subspace of the total Hamiltonian. The time dependent reduced density matrix for the $n^{\rm th}$ spin in the chain is analytically attainable and takes the convenient form~\cite{Campbell2011PRA},
\begin{equation}
\rho^{(n)}(t) = \left( \begin{array}{cc}
\rho_{0,0}^{(1)}(0) + \rho_{1,1}^{(1)}(0) \left( 1 - |f_n(t)|^2 \right) & \rho_{0,1}^{(1)}(0) f_n(t) \\
\rho_{1,0}^{(1)}(0) f_n^*(t) & \rho_{1,1}^{(1)}(0) |f_n(t)|^2
\end{array} \right),
\label{eq:dens_of_nth_qubit}
\end{equation}
where $\rho_{i,j}\!=\!\bra{i} \rho \ket{j}$ are the elements of the density matrix. Evaluating Eq.~\eqref{eq:dens_of_nth_qubit} requires determining the function $f_n(t) \!=\! _1\bra{\underline{\mathbf{1}}} e^{-i\mathcal{H}t} \ket{\underline{\mathbf{1}}}_n$ which corresponds to the transition amplitude for the $n^{\rm th}$ site in the chain and depends on the specific choice of coupling strengths. 

In the two limits we can determine this function analytically finding it is given by,
\begin{equation}
    f_n(t) = \left(\frac{1}{2}\right)^{\frac{N-1}{2}}\sqrt{\binom{N-1}{n-1}} e^{-iNBt}\sum_{k=1}^{N}K_{k-1}(n-1)e^{-iE_kt},
\end{equation}
for PST couplings and 
\begin{equation}
f_n(t) = \frac{2}{N+1} e^{-iNBt}\sum_{k=1}^{N} \sin\left(\frac{k\pi}{N+1}\right) \sin\left(\frac{k\pi n}{N+1}\right) e^{-i E_k t},
\label{eq:TransitionAmplitude}
\end{equation}
for the uniform coupling (see Appendices~\ref{CA} and \ref{XX} which also include the thermodynamic limit).

\begin{figure*}
    (a) \hskip0.33\linewidth (b) \hskip0.33\linewidth (c)
    \includegraphics[width=1\linewidth]{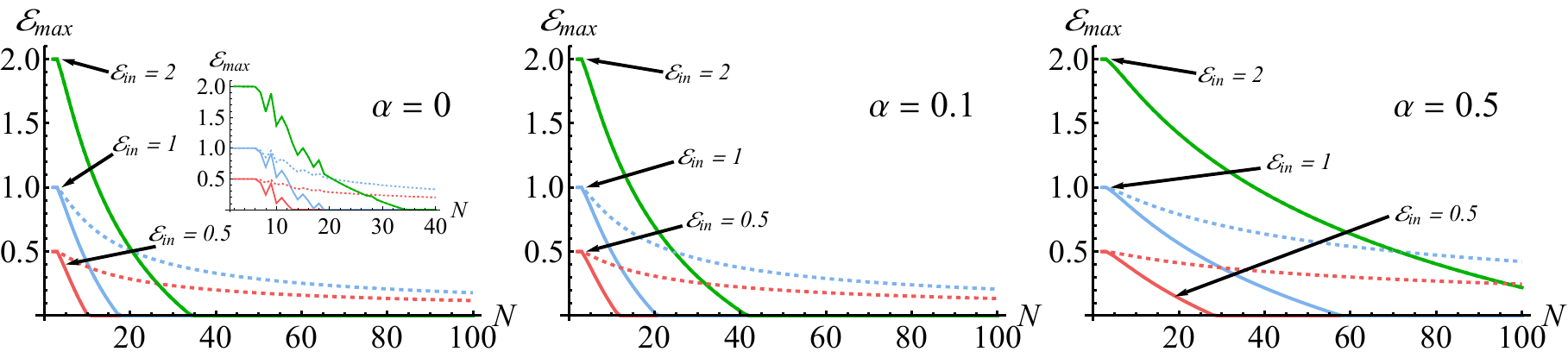}
    \caption{$\varepsilon_{max}$ of the last qubit in the chain, as a function of $N$, evaluated at the first reflection time, Eq.~\eqref{eq:ReflectTime}. The inset of the top panel is the same calculation but over a time interval $Jt\!\in\!(0,1000) \gg 2N$. The dashed curves correspond to initial states given by Eq.~\eqref{pureinitial} and the solid curves for Eq.~\eqref{mixedinitial}. Each pair of curves is initialized with the same $\varepsilon_{in}$ = $\{0.5,\,1,\,2\}$ corresponding to $\theta = \{\frac{\pi}{3},\,\frac{\pi}{2},\,\pi\}$ and $q=\{0.625,\,0.75,\,1\}$, with colours \{green, blue, red\} respectively and can be determined from Eqs. \eqref{eq:Ergcoh} and \eqref{eq:Ergmix}. Parameters: $J = 1,$ $B = 1$. Panels correspond to a different values of $\alpha$: (a)~0, (b)~0.1, and (c)~0.5.}
    \label{fig:Fig2}
\end{figure*}

From Eq.~\eqref{eq:dens_of_nth_qubit} it follows that for initial states given by Eq.~\eqref{mixedinitial} the resulting $n^{\rm th}$ site density matrix is always diagonal, while for initial states of the form Eq.~\eqref{pureinitial} the $n^{\rm th}$ site can acquire coherence. This already suggests that, with the notable exception of couplings that are engineered to achieve PST, storing the initial ergotropy ``classically", i.e. in populations, may generally be disadvantageous. In fact, notwithstanding that some energy will inevitably arrive at a given site after a sufficient time, in order for there to be extractable work from a diagonal state we require that $\rho^{(n)}_{1,1}\!>\!0.5$. As we will see in the proceeding subsections this places a fundamental bound on the maximum length of a uniform chain (or a chain with any non-PST set of couplings) that can achieve non-zero ergotropy transport. 

As our focus is on examining the locally extractable work that is transported along the chain, we assume the following protocol. At $t\!=\!0$ the interactions are suddenly quenched on and the whole system evolves for a time, $t$. The system evolves up to the first reflection time, \steve{i.e. the first time at which the transfer function $f_N(t)$ reaches a local maximum}, at which point we apply a swap operation between the last site of the chain and an auxiliary qubit with local Hamiltonian $H\!=\!-B\sigma_z$. We compute the ergotropy of this auxiliary system using Eq.~\eqref{eq:Ergotropy} which exactly corresponds to the ergotropy tranferred via the spin chain. It should be immediately clear that for $\alpha\!=\!1$ the PST couplings guarantee that whatever state is initially at site 1 will perfectly reach the end of the chain given a suitable amount of time, $Jt\!=\! \frac{\pi}{4\,G_N}N$ \cite{PSTQCA}, (and in fact this state will continue to bounce back and forth along the chain precisely localising at either end in a perfectly periodic manner). Thus, the strict limit of $\alpha\!=\!1$ guarantees the perfect shuttling of ergotropy, or in fact any quantity encoded in the initial state of the first site. However, this necessitates that the precise distribution of PST couplings is realised. We will therefore focus on $\alpha\!<\!1$, and in particular the uniform coupling limit $\alpha\!=\!0$, which will provide a lower bound for the ergotropy transport facilitated by our model. Furthermore, this focus will allow us to demonstrate that, in general, coherently endowed ergotropy can be more efficiently transported. 

\subsection{Ergotropy transport for non-PST coupling spin chains}
Using Eqs.~\eqref{eq:Ergotropy} and \eqref{eq:dens_of_nth_qubit}, we can readily determine the ergotropy of the $n^{\rm{th}}$ site in the chain. For ergotropy initially present in the pure state of the first site, Eq.~\eqref{pureinitial} we find,
\begin{equation}
\begin{split}
\varepsilon_{coh}(t) = {} & 2|f_n(t)|^2\sin^2\left(\frac{\theta}{2}\right) - \\
                         &  1 + \sqrt{1+4\sin^4\left(\frac{\theta}{2}\right)|f_n(t)|^2\left(|f_n(t)|^2-1\right)},
\end{split}
\label{eq:Ergcoh}
\end{equation}
while for ergotropy initially stored in the populations of the first site, Eq.~\eqref{mixedinitial}, we have,
\begin{equation}
    \varepsilon_{mix}(t) =\,2\left[2q\,|f_n(t)|^2 - 1\right], \quad q\,|f_n(t)|^2 >\frac{1}{2}.
    \label{eq:Ergmix}
\end{equation}
where the constraint follows since ergotropy is only non-zero for an active state. We begin by examining how effectively the chain is able to transport the extractable work for various chain lengths and amounts of initially stored ergotropy. For simplicity we focus on the first reflection time which (generally) corresponds to the maximum ergotropy transport, 
\begin{equation}
J t \!\approx\! T(\alpha,N)\! =\! \left[\frac{\pi}{4\, G_N}\alpha^2 + \frac{\pi}{6}\left(1-\alpha^2\right)\right]{N} 
\label{eq:ReflectTime}
\end{equation}
and determine the value of ergotropy that can be extracted from the end of the chain at site $N$ at this time, which we denote as $\varepsilon_{max}$. Eq.~\eqref{eq:ReflectTime} in the PST limit is known~\cite{PSTQCA}, and we numerically determine the quadratic form in $\alpha$ that interpolates between the two limits. \steve{We note that from the preceding, it follows that other sites in the chain may host extractable work during the dynamics. For initial states of the form Eq.~\eqref{mixedinitial} we find that the conditions necessary for most of the sites in the bulk of the chain to become active are not met, i.e. they do not achieve a population inversion. In contrast, initialising the chain in Eq.~\eqref{pureinitial} we find that the presence of coherence at a given site leads to the possibility of ergotropy being dynamically present. Nevertheless, in both cases, it is the hard-boundary that causes a ``refocussing" of the ergotropy at the last site of the chain and will be our primary focus.}

In  Fig.~\ref{fig:Fig2}(a) we show the behavior for various values of initial ergotropy, $\varepsilon_{in}$. Dashed curves corresponding to states where the extractable work is initially in the coherences, Eq.~\eqref{pureinitial}, while solid lines are for instances when it is initially in the population inverted state, Eq.~\eqref{mixedinitial}. For an initial state with maximal ergotropy on the first site, clearly there is no difference in the behaviour of the ergotropy since both initial states coincide. This highlights that it is not the purity of the state that governs the ability for ergotropy to be transported but rather the presence of coherences in the energy eigenbasis that are most relevant.  

Fig.~\ref{fig:Fig2}(a) demonstrates that for a mixed state with a given $\varepsilon_{in}$ there exists a strict chain-length cut-off beyond which no more work can be locally extracted. In contrast, for coherence endowed ergotropy, $\varepsilon_{max}$ scales to leading order as $\sim N^{-2/3}$. This scaling stems from the transition amplitude function in Eq.~\eqref{eq:TransitionAmplitude}, which for a uniformly coupled chain scales as $ f_N(T) \sim N^{-1/3}$~\cite{QcommReview} for large $N$ and sufficiently long times. This scaling can be seen by examining the thermodynamic limit of Eq.~\eqref{eq:TransitionAmplitude}, which we derive in Appendix~\ref{XX} and is discussed further in Ref.~\cite{QcommReview}. From Eq.~\eqref{eq:Ergcoh}, we see that $\varepsilon_{coh}(T)$, to leading order, scales as $|f_N(T)|^2 \!\propto\! N^{-2/3} $. Thus, for a finite system with coherences, ergotropy will reach the end of the chain given sufficient time. In panels (b) and (c) we show \steve{that more ergotropy can be faithfully transported} as we tune $\alpha$. For mixed initial states the maximum length of chain length for which ergotropy can still be transported increases and, as was the case for uniform couplings, the larger the amount of initial ergotropy present, the further it can be transported. For ergotropy initially endowed in coherences, increasing $\alpha$ has a similar impact insofar as the efficiency of the transport improves with increasing $\alpha$. For the moderate size systems shown here we see that the more initial ergotropy inputted corresponds to a larger amount reaching the end of the chain. While this is true in absolute terms for chains up to $N\sim100$ as shown, we remark that the dashed curves can cross for sufficiently large chains. Furthermore, if we consider the relative ergotropy transported, i.e. the efficiency $\varepsilon_{max}/\varepsilon_{in}$, a lower amount of initial ergotropy tends to be more faithfully transferred. This demonstrates that it is more efficient to transfer extractable work across long chains via coherences; the only exception to this being the strict limit of $\alpha\!=\!1$ where PST is achieved and therefore the ergotropy will be faithfully transferred regardless of how it is initially encoded.

In the inset in the left panel of Fig.~\ref{fig:Fig2}(a) we relax the assumption that the first reflection time corresponds to the maximum transport. We consider a large time window and numerically determine $\varepsilon_{max}$ at the last site of the chain for a substantially enlarged time window, $J t\!\in\!\left[0, 1000 \right]$. While the overall trend is consistent with the results shown in the main panel, we see that for specific chain lengths there can be instances of more extractable work reaching the end of the chain. This is a consequence of interference effects over long times, as the energy is shuttled back and forth along the chain, constructive interference can occur. However, these instances are highly parameter specific, in particular with the system size.

\begin{figure}[b]
    \centering
    \includegraphics[width=1  \linewidth]{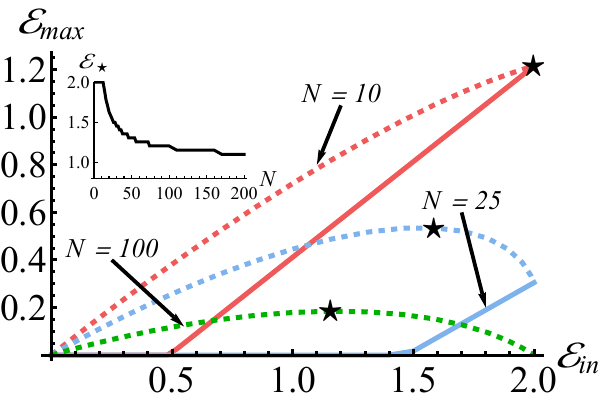}
    \caption{$\varepsilon_{max}$ endowed in coherences (dashed curves) and populations (solid curves) of the last qubit in the chain evaluated at $Jt$ given by Eq.~\eqref{eq:ReflectTime}, as a function of $\varepsilon_{in}$. We show three system sizes, $N=\{10,25,100\}$. Parameters: $\alpha = 0, J=1, B=1$. {\it Inset:} \steve{The value of $\varepsilon_{in}$ that corresponds to the peak value of $\varepsilon_{max}$, as a function of chain size $N$. The corresponding value is marked with a star for the system sizes shown in the main panel.}}
    \label{fig:Fig3}
\end{figure} 

\begin{figure}[t]
   (a) \hskip0.45\columnwidth (b)
    \includegraphics[width=1\linewidth]{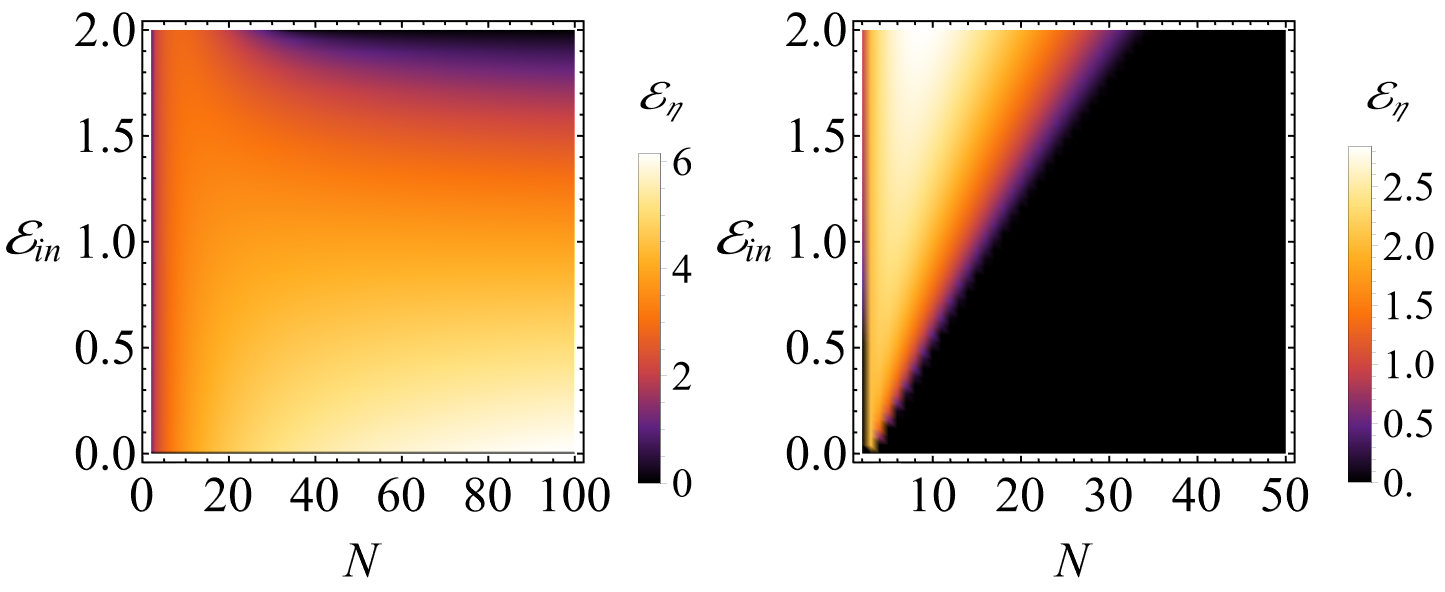}
    \caption{$\varepsilon_{\eta}$, calculated at $Jt$ given by Eq.~\eqref{eq:ReflectTime} as a function of $\varepsilon_{in}$ endowed in coherences (a) and populations (b) and $N$. Parameters: $\alpha = 0, J=1, B=1$ for both panels.}
    \label{fig:enctl}
\end{figure}

We further consider the efficiency of the ergotropy transport in Fig.~\ref{fig:Fig3} where we show the extractable work at the last site of the chain, $\varepsilon_{max}$, as a function of the initial ergotropy, $\varepsilon_{in}$, for the uniform coupling case. The dashed curves show the behavior for Eq.~\eqref{pureinitial}, i.e. an active pure state. For small chains, $N\sim 10$, increasing $\varepsilon_{in}$ leads to a greater amount of ergotropy transported. However, for larger chains we see this no longer holds. As $N$ is increased, we find the optimal initial state to ensure maximum transport of useful work tends to an input state that is an equal coherent superposition on the first site. \steve{Thus, for initial states of the form Eq.~\eqref{pureinitial} and finite chains, there is a non-trivial interplay between the role of coherence in facilitating ergotropy transport over longer chains and the benefits of having more initial ergotropy which necessitates a population inversion.} In contrast, for a mixed state, the higher the population of the excited state, the more ergotropy will reach the end of the chain, however, again we find that there is a strict cut off in chain length above which no useful work can be transferred. In the inset of Fig.~\ref{fig:Fig3} we plot the \steve{the value of $\varepsilon_{in}$, labelled $\varepsilon_{\star}$, that corresponds to the peak of each of the dashed curves as a function of chain length, $N$, from which we see that in the thermodynamic limit it is most favourable to initialise the state of the first qubit in an equal superposition.} We  show this analytically using the thermodynamic limit expression of Eq.~\eqref{eq:TransitionAmplitude} in Appendix~\ref{XX}. 

Since $\varepsilon_{max}$ scales to leading order with $N^{-2/3}$, finally we consider the rescaled efficiency 
\begin{equation}
    \varepsilon_\eta = \frac{\varepsilon_\text{max}}{\varepsilon_\text{in}}\; N^{2/3},
    \label{eq:Ergfom}
\end{equation}
which allows us to more clearly assess how the manner in which the useful work is initially encoded impacts ergotropy transport. In Fig.~\ref{fig:enctl}(a) we show Eq.~\eqref{eq:Ergfom} for coherently endowed initial ergotropy. We see that lower values of $\varepsilon_{in}$ are more efficiently transported. As the chain length increases, we see the optimal input states are initial states with less than the maximum possible $\varepsilon_{in}$. In panel (b) we show the same quantity this time for when ergotropy is initially encoded in the populations. For this initial state, we clearly see the cutoff in chain length scaling linearly with $\varepsilon_{in}$, with initial states with maximal ergotropy providing the greatest transport efficiency. 

\subsection{Robustness to disorder}
While the previous sections considered non-PST couplings, with a particular focus on uniform couplings, we now focus on PST. We will once again compare the ability for the chain to shuttle ergotropy depending on if the how initial extractable work is encoded. We will consider the PST coupling regime ($\alpha=1$) and disorder in the coupling strength $\mathcal{J}_\alpha$. The disorder is drawn from a uniform distribution in the range $[-\Delta,+\Delta]$~\cite{Ronke2011PRA, IreneDisorder2,magnonTransport,Coopmans2022}, where $\Delta$ quantifies the disorder strength. \steve{We remark that for weak disorder strengths, the impact of the noise is dictated by the variance, rather than the particular choice, of the distribution~\cite{DisorderPRX}. Therefore, we choose a uniform distribution as it effectively represents a worst case scenario as the strength of the disorder is increased~\cite{IreneDisorder2}.}

\begin{figure}[t]    \includegraphics[width=0.88\linewidth]{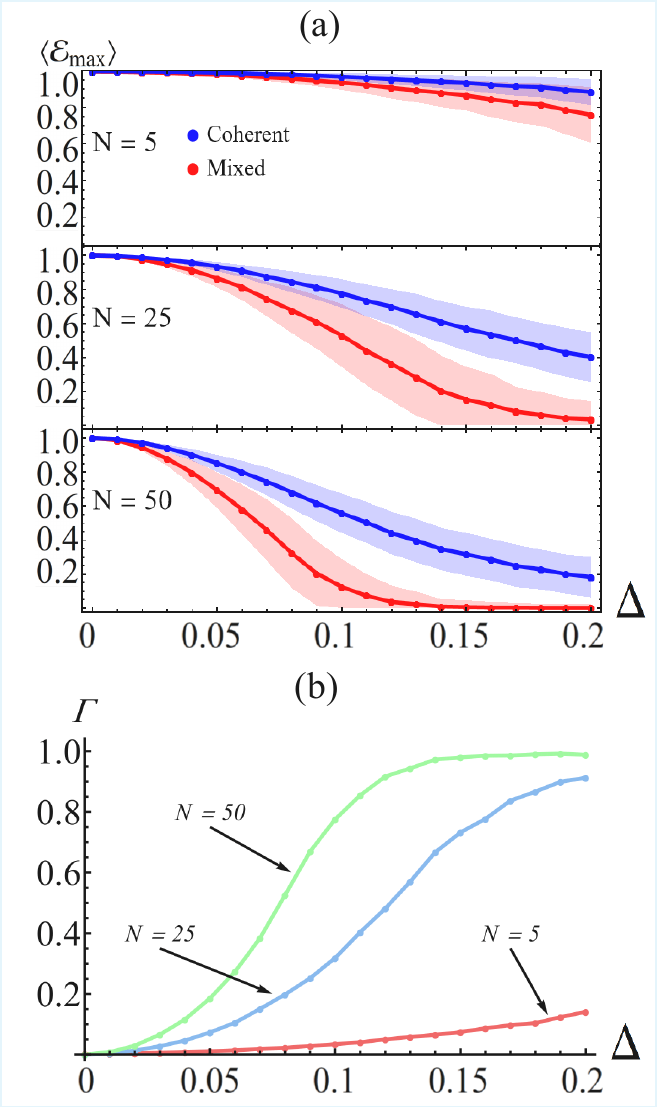}
    \caption{(a) $\varepsilon_{max}$ averaged over $1000$ realisations of disorder amplitude $\Delta$ evaluated at the first reflection time, Eq.~\eqref{eq:ReflectTime} with $\alpha\!=\!1$. The curves from top to bottom in both panels correspond to $N=\{5,25,50\}$. The upper, blue and lower, red curves correspond to coherent and mixed states, respectively.  The shaded region denotes one standard deviation from the mean. (b) Figure of merit from Eq.~\eqref{eq:gamma} as a function of $\Delta$ for various $N$. This is computed from the data in (a).}
    \label{fig:Disorder1}
\end{figure}

In Fig.~\ref{fig:Disorder1}(a) we show $\langle \varepsilon_{max}\rangle $ obtained at the first reflection point where $\langle \cdot \rangle$ denotes the ensemble average over $1000$ disorder realizations. We can see clearly see that for all system sizes and disorder strengths, the coherent superposition outperforms the mixed state. The superposition case also exhibits smaller deviations from the mean value, as captured by the size of the shaded regions which correspond to one standard deviation. We remark that the initial quadratic drop in ergotropy can be understood and quantified using either master equations for the ensemble average state~\cite{DisorderPRX,kiely2021} or the use of perturbation theory~\cite{poggi2024}.

This gap in performance between the two types of ergotropy storage, can be further highlighted by using the following figure of merit,
\begin{equation}
\label{eq:gamma}
\Gamma=\frac{\langle\varepsilon_{max}^{coh}\rangle-\langle \varepsilon_{max}^{mix}\rangle}{\langle\varepsilon_{max}^{coh}\rangle}.  
\end{equation}
This acts as a quantifier of the efficiency ergotropy transport in the presence of disorder for quantum versus classical ergotropy encoding. For states with the same amount of initial ergotropy, a $\Gamma\! >\! 0$ indicates that pure states with coherence outperform classical mixtures. Fig.~\ref{fig:Disorder1}(b) demonstrates that pure states with coherences in the context of ergotropy transfer are more robust to disorder compared to mixed states, $\Gamma\geq0$, for all the values of $N$ and $\Delta$ considered. 

\begin{figure*}[t]
\includegraphics[width=1\linewidth]{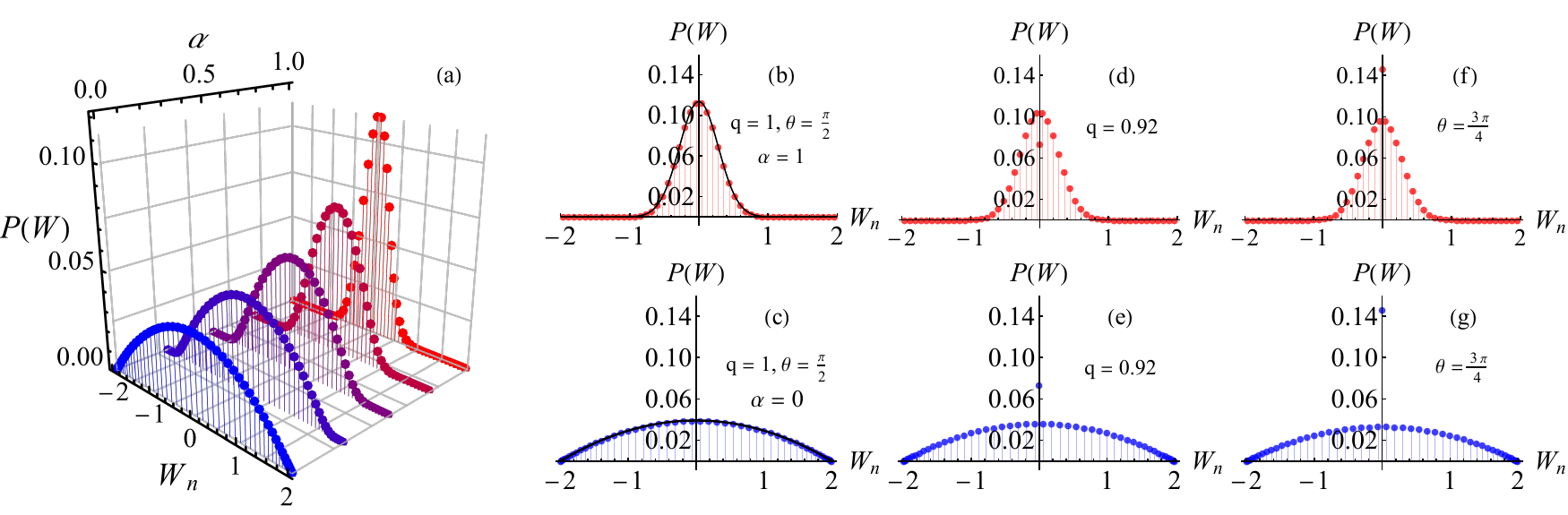}
\caption{(a) Work distribution, $P(W)$, resulting from suddenly quenching on the interactions in the chain interpolating between uniform (leftmost, blue, $\alpha\!=\!0$) and PST (rightmost, red, $\alpha\!=\!1$) regimes. (b) and (c) Show the two limits together with the probability density functions, Eq.~\eqref{eq:pdfCA} and Eq.~\eqref{eq:pdfXX}, shown by the continuous black lines. For panels (a-c) we assume the first site is in a maximally charged state, i.e. $\ket{1}$ corresponding to $\theta\!=\!\pi$ and $q\!=\!1$. (d-g) Work distributions corresponding to states with the same initial ergotropy present in the populations (d,e) or the coherences (f,g) of the first site. The top-row corresponds to PST couplings, $\alpha\!=\!1$ and the bottom row for uniform couplings $\alpha\!=\!0$. Panels (d) and (e) correspond to an active mixed initial state with $q\!=\!0.92$. Panels (f) and (g) correspond to a pure, coherent initial state with $\theta\!=\!\frac{3\pi}{4}$. All panels assume $J\! = \!1$, $B\!=\!1$, and $N\!=\!50$.}
    \label{fig:WD_XX}
\end{figure*}

\section{Thermodynamic Cost of Quench Protocol}
\label{sec:WorkDistResults}
In the preceding sections, we have examined the transport of extractable work, which as we have shown has subtle differences with the transport of population. However, there is a work cost associated with the sudden quench of the interactions needed for this transport to occur. In the next section, we will quantify this thermodynamic cost using the well known quantum work statistics~\cite{WorkNotObservable, Silva2008, Santini2023, Zawadzki2020, NonGaussworkStat, EntQWD, apollaro2015work}.

We assume an initial Hamiltonian $\mathcal{H}^i$ where the couplings are switched off (i.e. Eq.~\eqref{eq:Hxx} with $\mathcal{J}_\alpha\!=\!0$) and a final Hamiltonian $\mathcal{H}^f$ with the couplings on. Assuming the spectral decomposition for these Hamiltonians, i.e. $\mathcal{H}^{i,f}\!=\!\sum_m E_m^{i,f} \ketbra{E_m^{i,f}}{E_m^{i,f}}$, 
we can evaluate work involved in suddenly quenching on the interaction via the two point measurement (TPM) protocol~\cite{WorkNotObservable}. The TPM protocol provides us access to the quantum work distribution,
\begin{equation}
    \label{eq:WorkDist}
    P(W) = \sum_{m,k}^N p_m p_{k|m} \delta(W-W_{k,m}).
\end{equation}
For a given initial state, $\varrho$, the first measurement has probability $p_{m}\! =\! \bra{E_m^i}\varrho\ket{E_m^i}$. The conditional probability for the second measurement follows as $p_{k|m}\! =\! \bra{E_k^f}\varrho\ket{E_m^i}$, where we have already assumed the sudden quench limit where the interactions are switched on effectively instantaneously. The observed values of work correspond to energy differences from the two measurements, $W_{k,m} \!=\! E_k^f - E_m^i$. For the initial states considered in this work, Eqs.~\eqref{pureinitial} and \eqref{mixedinitial}, it follows that for $p_m$ there are at most only two possible non-zero values. From Eq.~\eqref{eq:WorkDist} the moments of the work distribution follow as $\langle W^n \rangle\! = \! \int dW W^nP(W)$, which are a commonly used summary statistic~\cite{EntQWD,NonGaussworkStat}. 

In Fig.~\ref{fig:WD_XX}(a) we fix the first site to be in a maximally active state, i.e. $q\!=\!1$ and/or $\theta\!=\!\pi$, and we examine how the work distribution changes as we interpolate between a uniform coupling and the PST couplings by varying $\alpha$ in Eq.~\eqref{eq:Hxx}. It is immediately clear that the resulting distributions are symmetric about the origin. Directly computing the first moment of the distribution confirms that the average work performed in switching on the interactions is $\langle W\rangle\!=\!0$, independently of the value of $\alpha$, i.e. regardless of whether the interactions will achieve PST or not. Thus, the first moment is not particularly informative regarding assessing the thermodynamics of communication through a spin chain, but simply implies that, on average, there is no work associated with switching on the channel. The full work statistics, however, clearly captures notable differences in the characteristics of the distribution as we tune between uniform and PST couplings. In particular, we see that when the system is engineered to achieve PST, the corresponding work distribution tends to a Gaussian while for the uniform coupling it is a parabola. It follows that the PST couplings have a smaller variance of work, i.e. fluctuations in the cost of switching on the interactions are smaller, and therefore such couplings can be considered more energetically stable.

As the eigenvalues and eigenstates, Eqs.~\eqref{eq:EigvalsCA}--\eqref{eq:EigvecsXX}, are exactly known in the two limits, we can analytically determine $P(W)$ for $q\!=\!1$ and $\theta\!=\!\pi$. In the PST limit with $\alpha \!=\! 1$ we find
\begin{equation}
\label{PW_PST}
    P(W) = \left( \frac{1}{2} \right)^{N-1}\sum_{k=1}^{N} \binom{N-1}{k-1} 
    \delta(W - W_k).
\end{equation}
where $W_k \!=\!  -\dfrac{2J}{N}[N - (2k - 1)] G_N $ for $k \geq 1$.
For large $N$ Eq.~\eqref{PW_PST} is well approximated by the probability density,
\begin{eqnarray}
    p(W) \approx \frac{1}{\sqrt{2\pi\,\mathrm{Var}(W)}} \exp\left[-\frac{W^2}{2\,\mathrm{Var}(W)}\right].
    \label{eq:pdfCA}
\end{eqnarray}
with variance $\mathrm{Var}(W)\! =\! 2\left(\frac{2J}{N}\right)^2(N-1)G_N^2$, see Appendix~\ref{CA} for details of the derivation.  
We remark that for $N$ finite $\langle W^n \rangle\neq 0$ for all even $n\!\geq\!2$, i.e. the distribution will have non-zero higher order even moments. However, these moments vanish as $N$ increases and the exact distribution, Eq.~\eqref{PW_PST}, becomes Gaussian. In the thermodynamic limit the distribution converges to a Dirac delta function at the origin. 

Turning to the uniform coupling limit, $\alpha\!=\!0$, the work distribution is given by,
\begin{equation}
    P(W) = \frac{2}{N+1}\sum_{ k=1}^{N}\sin^2\left(\frac{k\pi}{N+1}\right)\delta\left(W-W_{k}\right),
    \label{eq:WDxx}
\end{equation}
with $W_k \!=\! -2J\cos\left(\frac{k\pi}{N+1}\right)$ for $k \geq 1$. This distribution follows a parabola, as shown in Fig.~\ref{fig:WD_XX}(a) and we find it has a variance $\mathrm{Var}(W)\! =\! J^2$ independent of $N$. In the thermodynamic limit, we obtain the probability density function (see Appendix~\ref{XX}),
\begin{equation}
     p(W) = \frac{1}{2\pi J^2}\sqrt{4J^2-W^2}.
     \label{eq:pdfXX}
\end{equation}
From Fig.~\ref{fig:WD_XX}(a) we see that as $\alpha$ is varied, the distribution smoothly interpolates between the two regimes, becoming progressively more Gaussian as we approach the PST couplings.

In Fig.~\ref{fig:WD_XX}(b-g) we show the exact work distributions for various initial states with fixed chain length $N\!=\!50$. The top row corresponds to PST couplings while the bottom row shows the results for uniform couplings. In panels (b) and (c) we again show the behavior for the same maximally charged initial state as considered in (a). We also show the corresponding probability density functions, Eq.~\eqref{eq:pdfCA} and Eq.~\eqref{eq:pdfXX}, respectively showing that an excellent agreement is achieved for moderate sized chains. 

In Figs.~\ref{fig:WD_XX}(d)-(g) we see that the main features discussed above persist regardless of the specific initial state of the system. Different initial states principally impact the probability of doing zero work, $P(0)$. We see that regardless of whether the initial ergotropy is present in populations, panels (d) and (e) or coherences, panels (f) and (g), the first measurement of the TPM results in a non-zero probability of collapsing the first site to its ground state. In this case, it follows that the chain's initial state after the first measurement becomes $\ket{\underline{\mathbf{0}}}$, which corresponds to the $k\!=\!0$ eigenstate of the Hamiltonian and is independent of $J_\alpha$. Therefore no work is performed in switching from $J_\alpha\!=\!0$ to $J_\alpha\!\neq \!0$. The magnitude of $P(0)$ is then wholly determined by the first site's $\ket{0}$ amplitude/population. It follows that the observed spike in $P(0)$ will generally be more prominent for the pure initial state since it requires only a moderate amount of coherence to store the same amount of ergotropy compared to the population inverted state. For the chosen parameters in Fig.~\ref{fig:WD_XX}(f)-(g) the pure state has a large amplitude in the $\ket{0}_1$ state and therefore leads to a higher probability for $P(0)$ (see panel (g)). In contrast, for the mixed initial state, Eq.~\eqref{mixedinitial}, we require $q\!>\!1/2$ and therefore the population of the local ground state, which directly determines $P(0)$, is smaller compared to the pure initial state.

It is worth commenting that the preceding analysis also applies more generally to assessing the thermodynamics of quantum communication protocols~\cite{deChiara2005PRA, Ronke2011PRA, kay2010perfect}. Our results demonstrate that at the level of the work done in switching on the necessary interactions, there are notable differences that arise depending on the precise distribution of couplings employed. Interestingly, we see that ideal communication channels exhibit smaller fluctuations in the work cost of turning on the couplings. 

\section{Conclusions}
We have examined the utility of one-dimensional spin chains as a means to transfer extractable work. We considered a two-level system acting as a battery, initalised in an active state, where the ergotropy encoded in either quantum coherence and/or via a population inversion. We have established that there is a clear quantum advantage to encoding extractable work in coherences. For uniform couplings, this encoding allows for more efficient transport over longer spin chains, while for couplings that achieve perfect state transfer (PST) we demonstrated that coherently endowed ergotropy is more robust to disorder. Furthermore, it is worth noting that our results also highlight the subtle distinction between the transport of other figures of merit, e.g. the quantum state, spin, or energy, compared to {\it useful} work. We have also assessed the work cost of quenching on the chain couplings, establishing that while for all excitation preserving interactions the average work cost is zero, PST-type couplings lead to work distributions with a lower variance and therefore are more stable to fluctuations. In addition to contributing to the growing interest in assessing the capabilities of quantum systems to act as energy storage devices, i.e. batteries, our work also provides insight into the thermodynamics of quantum networks and quantum communication channels. 

\acknowledgements
DM, ID’A, and SC are grateful to the Royal Society International Exchanges Scheme IES$\backslash$R2$\backslash$242072. DM acknowledges support from the UCD School of Physics Scholarship in Research and Teaching. AK acknowledges financial support of Taighde \'Eireann – Research Ireland under grant number 24/PATH-S/12701.

\appendix

\section{XX PST Coupling Analytics}
\label{CA}
Here we work in the single-excitation subspace, following the results of \cite{Albanese2004}. This allows us to obtain analytical expressions for the transition amplitude function and the work distribution when the XX model is tuned with PST coupling. The nearest-neighbor hopping Hamiltonian for a chain with position-dependent couplings is given by
\begin{equation}
\mathcal{H} = \frac{1}{2}\sum_{j=1}^{N-1} J_{j,j+1} \bigl( \sigma_j^x \sigma_{j+1}^x + \sigma_{j+1}^y \sigma_j^y \bigr).
\end{equation}
The PST coupling is defined by
\begin{equation}
J_{j,j+1} = \frac{2J}{N}\,\sqrt{j\,(N-j)}G_N.
\end{equation}
When we apply \(\mathcal{H}\) to a single-particle state, the eigenvalue equation \(\mathcal{H}\,\ket{E_k} = E_k\,\ket{E_k}\) turns into a difference equation for the amplitudes \(\psi(j)\) of the state, $\ket{E_k}=\sum_j \psi(j) \ket{\underline{\mathbf{1}}}_j$, obtaining,
\begin{equation}
\begin{split}
\frac{2J}{N}\,\sqrt{(j-1)(N-j+1)}\,\psi(j-1)G_N \;+ \\
 \frac{2J}{N}\,\sqrt{(j+1)(N-j-1)}\,\psi(j+1)G_N 
= E\,\psi(j)
\end{split}
\label{eq:ThreeTerm}
\end{equation}
with the natural boundary conditions
$\psi(0) = \psi(N+1)=0.$
This recurrence relation can be shown to be equivalent to an eigenvalue equation for Krawtchouk polynomials~\cite{Christandl2017},
\begin{equation}
K_k(n-1; p, N-1)
= \sum_{i=0}^{k} (-1)^i \, \binom{n-1}{i} \, \binom{N  - n}{k - i} \left(\frac{p}{1-p}\right)^i,
\label{eq:Kpol}
\end{equation}
where in our case $p=\frac{1}{2}$, enforcing symmetry in the weight function, $w(j)$, see below, which is a necessary property in the Hamiltonians' structure to enable PST.  

The solutions to the difference equation, Eq.~\eqref{eq:ThreeTerm}, are given by,
\begin{equation}
\psi_k(j) = \sqrt{w(j)}\, K_k\Bigl(j; \tfrac{1}{2}, N\Bigr),\quad k=1,2,\dots,N,
\end{equation}
where \(K_k\left(j; \tfrac{1}{2}, N\right)
\) are the Krawtchouk polynomials with eigenstates $|E_k\rangle = \sum_{j=1}^{N} \psi_k(j) \ket{\underline{\mathbf{1}}}_j$. The weight function is \( w(j)=\binom{N-1}{j-1}\left(\frac{1}{2}\right)^{N-1},\)
which arises naturally from the orthogonality of these polynomials. With the appropriate normalisation, the eigenvalues are linear in $k$,
\begin{equation}
E_k = \frac{2J}{N}\left[N - (2k-1)\right]G_N,\quad k=1,2,\dots,N.
\end{equation}

We can obtain an expression for the transition amplitude function,
 $   f_n(t) \!=\! \bra{\underline{\mathbf{1}}}_1 e^{-i\mathcal{H}t} \ket{\underline{\mathbf{1}}}_n $
by directly substituting the eigenstates and eigenvalues in  $\mathcal{H}\!=\!\sum_k E_k \ketbra{E_k}{E_k}$, to obtain,
\begin{eqnarray}
f_n(t) &=&\left(\frac{1}{2}\right)^{\frac{N-1}{2}} \sqrt{\binom{N-1}{0}\binom{N-1}{n-1}}\sum_{k=1}^{N}K_{k-1}(0)K_{k-1}(n-1)e^{-iE_kt} \nonumber \\
&=& \left(\frac{1}{2}\right)^{\frac{N-1}{2}} \sqrt{\binom{N-1}{n-1}}\sum_{k=1}^{N}K_{k-1}(n-1)e^{-iE_kt}.
\end{eqnarray}

Using the orthogonality relations and that $K_{k-1}(0)=K_{k-1}(1) = 1,\;\forall\; k$, calculating $\left|\langle E_k | \underline{\mathbf{1}} \rangle_1\right|^2$, we can obtain the work distribution of a sudden quench of the PST coupling in the XX model as,
\begin{equation}
    P(W) =\left( \frac{1}{2} \right)^{N-1} \sum_{k=1}^{N} \binom{N-1}{k-1} 
    \delta(W - W_k).
    \label{eq:WD_CA_derive}
\end{equation}
This is binomially distributed in $(k-1)$.  Considering that $W_k = E_k$, a linear transformation in $(k-1)$  rescales the unit variance of the binomial while maintaining a zero mean. These results extend trivially when introducing a uniform magnetic field term at each site, introducing a global phase, with no contribution to the work distribution. We identify a random variable, $X = k-1$, such that 
\begin{eqnarray}
    W_X = -\frac{2J}{N}(N-2X-1)G_N.
    \label{eq:WX}
\end{eqnarray}
To see how Eq.\eqref{eq:WD_CA_derive} behaves in the thermodynamic limit we first note the mean and variance of the binomial distribution in $X$,
\begin{equation}
\mu  = \frac{N-1}{2}, \qquad \sigma^2 = \frac{N-1}{4}.
\end{equation}

By the de Moivre–Laplace theorem, the binomial distribution converges to a normal distribution under proper centering and scaling. More specifically for large $N$,
\begin{equation}
\binom{N-1}{k-1} \left( \frac{1}{2} \right)^{N-1} 
\approx \frac{1}{\sqrt{2\pi \sigma^2}} \exp\left[ -\frac{\tilde{X}^2}{2} \right].
\label{eq:BinomWD}
\end{equation}
where we have defined the standardised random variable
\begin{equation}
\tilde{X}=\frac{(k-1) - \mu}{\sqrt{\sigma^2}}.
\label{eq:DRV}
\end{equation}

From Eq.\eqref{eq:WX}, calculating the variance of $W_X$ in the thermodynamic limit and using that a constant shift does not alter the variance, we obtain,
\begin{eqnarray}
    \mathrm{Var}(W_X) = \left[2\left(\frac{2J}{N}\right)G_N\right]^2\mathrm{Var}(X) = \left(\frac{2J}{N}\right)^2(N-1)G_N^2,
\end{eqnarray}
with the overall $N$ dependence,
\begin{eqnarray}
    \mathrm{Var}(W_X) \sim  \left(\frac{1}{N}\right)^2N = \mathcal{O}(N^{-1}) \to 0 \quad \text{as } N \to \infty.
\end{eqnarray}

Eq.\eqref{eq:WD_CA_derive} becomes a sequence of normalised probability distributions with vanishing variance which converges to a Dirac delta function centred at the mean,

\begin{eqnarray}
    \delta(x) = \lim_{\sigma \to 0} \frac{1}{\sqrt{2\pi}\,\sigma} \exp\left(-\frac{x^2}{2\sigma^2}\right).
\end{eqnarray}

Putting it all together, we get the probability density function for large N,
\begin{eqnarray}
    p(W)dW \approx \frac{1}{\sqrt{2\pi\,\mathrm{Var}(W)}} \exp\left[-\frac{W^2}{2\,\mathrm{Var}(W)}\right]dW 
\end{eqnarray}
and in the thermodynamic limit it converges to the Dirac delta function $p(W)dW = \delta(W)dW$.

\section{XX Uniform Coupling Analytics}

\label{XX}

\subsection{Transition probabilities}
To derive Eq.\eqref{eq:TransitionAmplitude}, we first write the expression as,
\begin{equation}
   f_n(t) \!=\! \bra{\underline{\mathbf{1}}}_1 e^{-i\mathcal{H}t} \ket{\underline{\mathbf{1}}}_n.
   \label{eq:fnapp}
\end{equation}
Using the Jordan-Wigner transformations on Eq.~\eqref{eq:Hxx} in the uniform coupling limit ($\alpha=0$), we obtain the energy eigenstates to be,
\begin{equation}
    \ket{E_k}=\sqrt{\frac{2}{N+1}}\sum_{n=1}^N\sin\left( \frac{\pi k n}{N+1}\right)\ket{\underline{\mathbf{1}}}_n,
    \label{eq:EigvecsXXapp}
\end{equation}
with eigenvalues,
\begin{equation}
    E_k = -2J\cos\left({\frac{\pi k}{N+1}}\right)-(N-2)B.
    \label{eq:EigvalsXXapp}
\end{equation}
Rearranging this,
\begin{equation}
    \ket{\underline{\mathbf{1}}}_n=\sqrt{\frac{2}{N+1}}\sum_{k=1}^N\sin\left( \frac{\pi k n}{N+1}\right)\ket{E_k},
    \label{eq:EigstatesXXapp}
\end{equation}
we can substitute $\ket{\underline{\mathbf{1}}}_n$ and $\ket{\underline{\mathbf{1}}}_1$ into Eq.~\eqref{eq:fnapp} to get the desired expression,
\begin{equation}
f_n(t) = \frac{2}{N+1} \sum_{k=1}^{N} \sin\left(\frac{k\pi}{N+1}\right) \sin\left(\frac{k\pi n}{N+1}\right) e^{it \left[(N-2)B + 2J\cos\frac{k\pi}{N+1} \right]}.
\label{eq:fn}
\end{equation}
Note that when including the zero-excitation subspace a global phase, $e^{-iNBt}$, picked up.

\begin{figure}[t]
    \centering
    \includegraphics[width=0.8\linewidth]{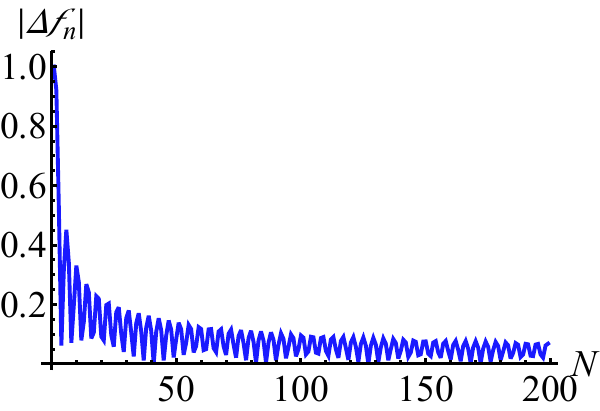}
    \caption{Difference in $f_N\left(T\right)$ between the discrete and continuous expressions.}
    \label{fig:Fig7}
\end{figure}
In the thermodynamic limit, this becomes
\begin{eqnarray}
    f_n(t) \xrightarrow{N \to \infty} \frac{2}{\pi} e^{i (N-2)B t} \int_0^\pi dx \, \sin(x) \sin(n x) e^{2iJt \cos(x)}.
\end{eqnarray}
To simplify this, we employ the Jacobi-Anger expansion,
\begin{equation}
e^{i z \cos \theta} = J_0(z) + 2 \sum_{l=1}^\infty i^l J_l(z) \cos(l \theta).
\end{equation}
This gives
\begin{align}
f_n(t) = \frac{2}{\pi} e^{i A t} \Bigg[ 
& J_0(2Jt) \int_0^\pi dx \, \sin(x) \sin(n x) \nonumber \\
& + 2 \sum_{l=1}^\infty i^l J_l(2Jt) \int_0^\pi dx \, \sin(x) \sin(n x) \cos(l x) 
\Bigg],
\end{align}
where $A=(N-2)B$.The first integral is a straightforward orthogonality relation,
\begin{equation}
\int_0^\pi dx \, \sin(x) \sin(n x) = 
\delta_{1,n}\frac{\pi}{2}.
\end{equation}
The second integral can be computed as
\begin{align}
  \frac{1}{2} \int_0^\pi dx \, \sin(x) \left[ \sin((n{+}l)x) + \sin((n{-}l)x) \right] \notag \\
  = \frac{1}{2} \cdot \frac{\pi}{2} \left( \delta_{1,n+l} + \delta_{1,n-l} \right) \nonumber
\end{align}
resulting in,
\begin{eqnarray}
f_n(t) &=& \frac{2}{\pi} e^{i A t} \left[ \frac{\pi}{2} J_0(2Jt) \delta_{1,n} + \sum_{l=1}^\infty i^l J_l(2Jt) \cdot \frac{\pi}{2} \left( \delta_{1,n+l} + \delta_{1,n-l} \right) \right] \nonumber \\
&=& e^{i A t} \left[ J_0(2Jt) \delta_{1,n} + i^{n-1} J_{n-1}(2Jt) \right],
\label{eq:fB} 
\end{eqnarray}
When the initial state is in a coherent superposition, $|f_n(T)|^2$, with $T$ from Eq.~\eqref{eq:ReflectTime}, gains a factor of $\mathrm{sin}^4(\frac{\theta}{2})$. Noting that the Bessel function of the first kind, $J_l$ is less than $1$ we can then see that for $n=N$, $|f_N(T)|^2$ is maximum when $\mathrm{sin}^4(\frac{\theta}{2})=1$, corresponding to $\theta=\pi$, i.e. an equal superposition. 
In Fig.~\ref{fig:Fig7} we plot the absolute value of the difference in the discrete and continuum expressions for $f_N\left(T\right)$.

\subsection{Work distribution}
The work distribution can also be derived using this procedure. The transition probabilities, $p_{k|m}$, reduce to $p_{k|1}$, which using the eigenstates, Eq.~\eqref{eq:EigvecsXXapp} takes the form,
\begin{equation}
    p_{k|1} = \fabs{\langle E_k | \underline{\mathbf{1}} \rangle_1}^2=\frac{2}{N+1}\sin^2\left(\frac{k\pi}{N+1}\right).
\end{equation}

Taking Eq.\eqref{eq:WDxx} in the thermodynamic limit ($N \to \infty$), the discrete values $W_k$ become dense in the interval $[-2J, 2J]$. As before we introduce a continuous variable $x = \frac{k\pi}{N+1}$, such that $W = -2J\cos(x)$. Then, $dx = \frac{\pi}{N+1} dk$, and $dW = 2J\sin(x) dx = 2J\sin(x) \frac{\pi}{N+1} dk$. Thus, $dk = \frac{N+1}{2J\pi\sin(x)} dW$.
The probability associated with a single $k$ is $p_k = \frac{2}{N+1}\sin^2(x)$. The probability density function $p(W)$ is obtained by considering the probability of $W$ falling in a small interval $[W, W + dW]$, which corresponds to a sum over the probabilities of the $k$ values in that interval,
\begin{equation}
p(W) dW \approx p_k dk = \frac{\sin(x)}{\pi J} dW 
\end{equation}
From $W = -2J\cos(x)$, we have $\cos(x) = -\frac{W}{2J}$. Using the identity $\sin^2(x) = 1 - \cos^2(x)$, we get,
\begin{equation}
\sin(x)  = \tfrac{1}{2J}\sqrt{4J^2 - W^2}.
\end{equation}
Substituting this back into the expression for $p(W) dW$, we get,
\begin{equation}
p(W) dW = \frac{\sqrt{4J^2 - W^2}}{2\pi J^2} dW.
\label{eq:PDFxxAPP}
\end{equation}
for $-2J \leq W \leq 2J$ and $p(W, J) = 0$ otherwise. While previous works have studied work statistics in an XX spin chain, for example \cite{Moriya2019}, the analytic expression for the probability density function in Eq.\eqref{eq:PDFxxAPP} has not been reported. This distribution has a zero first moment of work and a second moment $\langle W^2 \rangle = J^2$.

\bibliography{bibfile}

@article{IreneDisorder2, 
volume={5}, 
url={http://dx.doi.org/10.1002/qute.202200013}, DOI={10.1002/qute.202200013}, 
journal={Advanced Quantum Technologies}, 
volume = {5},
pages= {2200013},
author={Alsulami, A H. and D’Amico, I and Estarellas, M P. and Spiller, T P.}, 
year={2022}
}

@article{DisorderPRX,
  title = {Effective Dynamics of Disordered Quantum Systems},
  author = {Kropf, Chahan M. and Gneiting, Clemens and Buchleitner, Andreas},
  journal = {Phys. Rev. X},
  volume = {6},
  issue = {3},
  pages = {031023},
  numpages = {22},
  year = {2016},
  month = {Aug},
  publisher = {American Physical Society},
  doi = {10.1103/PhysRevX.6.031023},
  url = {https://link.aps.org/doi/10.1103/PhysRevX.6.031023}
}

@article{magnonTransport,
  author = {Kiely, A. and Campbell, S.},
  journal = {New J. Phys.},
  volume = {23},
  year = {2021},
  url = {https://api.semanticscholar.org/CorpusID:231592737}
}

@article{QcommReview,
  author = {Bose, S.},
  journal = {Contemp. Phys.},
  volume = {48},
  number = {1},
  pages = {13–30},
  year = {2007},
  month = {Jan},
  doi = {10.1080/00107510701342313},
  url = {http://dx.doi.org/10.1080/00107510701342313}
}

@article{QcommUnmod,
  author = {Bose, S.},
  journal = {Phys. Rev. Lett.},
  volume = {91},
  number = {20},
  pages = {207901},
  year = {2003},
  month = {Nov},
  doi = {10.1103/PhysRevLett.91.207901},
  url = {https://link.aps.org/doi/10.1103/PhysRevLett.91.207901}
}

@article{PSTQCA,
  author = {Christandl, M. and Datta, N. and Ekert, A. and Landahl, A. J.},
  journal = {Phys. Rev. Lett.},
  volume = {92},
  number = {18},
  pages = {187902},
  year = {2004},
  month = {May},
  doi = {10.1103/PhysRevLett.92.187902},
  url = {https://link.aps.org/doi/10.1103/PhysRevLett.92.187902}
}

@article{PSTarbQCA,
  author = {Christandl, M. and Datta, N. and Dorlas, T. C. and Ekert, A. and Kay, A. and Landahl, A. J.},
  journal = {Phys. Rev. A},
  volume = {71},
  number = {3},
  pages = {032312},
  year = {2005},
  month = {Mar},
  doi = {10.1103/PhysRevA.71.032312},
  url = {https://link.aps.org/doi/10.1103/PhysRevA.71.032312}
}

@article{Zawadzki2020,
  title = {Work-distribution quantumness and irreversibility when crossing a quantum phase transition in finite time},
  author = {Zawadzki, Krissia and Serra, Roberto M. and D'Amico, Irene},
  journal = {Phys. Rev. Res.},
  volume = {2},
  issue = {3},
  pages = {033167},
  numpages = {6},
  year = {2020},
  month = {Jul},
  publisher = {American Physical Society},
  doi = {10.1103/PhysRevResearch.2.033167},
  url = {https://link.aps.org/doi/10.1103/PhysRevResearch.2.033167}
}

@article{NonGaussworkStat,
  author = {Zawadzki, K. and Kiely, A. and Landi, G. T. and Campbell, S.},
  journal = {Phys. Rev. A},
  volume = {107},
  number = {1},
  pages = {012209},
  year = {2023},
  month = {Jan},
  doi = {10.1103/PhysRevA.107.012209},
  url = {https://link.aps.org/doi/10.1103/PhysRevA.107.012209}
}

@article{CorrLossErg,
  author = {Simon, R. P. A. and Anders, J. and Hovhannisyan, K. V.},
  journal = {Phys. Rev. Lett.},
  volume = {134},
  number = {1},
  pages = {010408},
  year = {2025},
  month = {Jan},
  doi = {10.1103/PhysRevLett.134.010408},
  url = {https://link.aps.org/doi/10.1103/PhysRevLett.134.010408}
}

@article{QCohErg,
  author = {Francica, G. and Binder, F. C. and Guarnieri, G. and Mitchison, M. T. and Goold, J. and Plastina, F.},
  journal = {Phys. Rev. Lett.},
  volume = {125},
  number = {18},
  pages = {180603},
  year = {2020},
  month = {Oct},
  doi = {10.1103/PhysRevLett.125.180603},
  url = {https://link.aps.org/doi/10.1103/PhysRevLett.125.180603}
}

@article{EntQWD,
  author = {Kiely, A. and O'Connor, E. and Fogarty, T. and Landi, G. T. and Campbell, S.},
  journal = {Phys. Rev. Res.},
  volume = {5},
  number = {2},
  pages = {L022010},
  year = {2023},
  month = {Apr},
  doi = {10.1103/PhysRevResearch.5.L022010},
  url = {https://link.aps.org/doi/10.1103/PhysRevResearch.5.L022010}
}

@article{NthRoot,
  author = {Fox, E. and Herrera, M. and Schmidt-Kaler, F. and D’Amico, I.},
  journal = {Entropy},
  volume = {26},
  number = {11},
  pages = {952},
  year = {2024},
  month = {Nov},
  doi = {10.3390/e26110952},
  url = {http://dx.doi.org/10.3390/e26110952}
}

@article{Allahverdyan2004,
  author = {Allahverdyan, A. E and Balian, R. and Nieuwenhuizen, Th. M.},
  journal = {Europhys. Lett.},
  volume = {67},
  number = {4},
  pages = {565–571},
  year = {2004},
  month = {Aug},
  doi = {10.1209/epl/i2004-10101-2},
  url = {http://dx.doi.org/10.1209/epl/i2004-10101-2}
}

@article{WorkNotObservable,
  author = {Talkner, P. and Lutz, E. and H\"anggi, P.},
  journal = {Phys. Rev. E},
  volume = {75},
  number = {5},
  pages = {050102},
  year = {2007},
  month = {May},
  doi = {10.1103/PhysRevE.75.050102},
  url = {https://link.aps.org/doi/10.1103/PhysRevE.75.050102}
}

@article{Albanese2004,
  author = {Albanese, C. and Christandl, M. and Datta, N. and Ekert, A.},
  journal = {Phys. Rev. Lett.},
  volume = {93},
  number = {23},
  pages = {230502},
  year = {2004},
  month = nov,
  doi = {10.1103/PhysRevLett.93.230502},
  url = {http://dx.doi.org/10.1103/PhysRevLett.93.230502}
}

@misc{inhomspinchain,
  author = {Bernard, P. A. and Parez, G. and Vinet, L.},
  eprint = {2411.09487},
  archivePrefix = {arXiv},
  primaryClass = {quant-ph},
  year = {2024},
  url = {https://arxiv.org/abs/2411.09487}
}

@article{Moriya2019,
  author    = {Moriya, H. and Nagao, R. and Sasamoto, T.},
  journal   = {J. Stat. Mech.},
  volume    = {2019},
  number    = {6},
  pages     = {063105},
  year      = {2019},
  month     = {jun},
  url       = {http://dx.doi.org/10.1088/1742-5468/ab1dd6}
}

@article{Kandel2021,
  author = {Y. P. Kandel and H. Qiao and S. Fallahi and G. C. Gardner and M. J. Manfra and J. M. Nichol},
  journal = {Nat. Commun.},
  volume = {12},
  number = {1},
  pages = {2154},
  year = {2021},
  publisher = {Springer},
  doi = {10.1038/s41467-021-22416-5}
}

@article{Qiao2021,
  author = {H. Qiao and Y. P. Kandel and S. Fallahi and G. C. Gardner and M. J. Manfra and X. Hu and J. M. Nichol},
  journal = {Phys. Rev. Lett.},
  volume = {126},
  number = {1},
  pages = {017701},
  year = {2021},
  doi = {10.1103/PhysRevLett.126.017701}
}

@article{kiely2021,
  author={Kiely, A.},
  journal={Europhys. Lett.},
  volume={134},
  number={1},
  pages={10001},
  year={2021},
  publisher={IOP Publishing},
url = {https://dx.doi.org/10.1209/0295-5075/134/10001}
}

@article{poggi2024,
  author={Poggi, P M and De Chiara, G and Campbell, S and Kiely, A},
  journal={Phys. Rev. Lett.},
  volume={132},
  number={19},
  pages={193801},
  year={2024},
  publisher={APS},
  url = {https://link.aps.org/doi/10.1103/PhysRevLett.132.193801}
}

@article{Li2025,
  title={},
  author={Li, L. and Zhao, S. and Xu, K. and Fan, H. and Zheng, D. and Xiang, Z.},
  journal={},
  year={},
  month={},
  eprint={2506.16881},
  archivePrefix={arXiv},
  primaryClass={}
}

@article{Hoang2024, 
title={}, 
volume={6}, 
ISSN={2643-1564}, 
url={http://dx.doi.org/10.1103/PhysRevResearch.6.013038}, 
number={1}, 
journal={Phys. Rev. Res.}, 
publisher={American Physical Society (APS)}, author={Hoang, D. T. and Metz, F. and Thomasen, A. and Anh-Tai, T. D. and Busch, T. and Fogarty, T.}, 
year={2024} }

@article{Coopmans2022,
  author = {Coopmans, L. and Campbell, S. and De Chiara, G. and Kiely, A.},
  journal = {Phys. Rev. Res.},
  volume = {4},
  issue = {4},
  pages = {043138},
  numpages = {10},
  year = {2022},
  month = {Nov},
  publisher = {American Physical Society},
  doi = {10.1103/PhysRevResearch.4.043138},
  url = {https://link.aps.org/doi/10.1103/PhysRevResearch.4.043138}
}

@article{Binder2019book,
  title={Thermodynamics in the Quantum Regime: Fundamental Aspects and New Directions},
  author={Binder, F. and Correa, L.A. and Gogolin, C. and Anders, J. and Adesso, G.},
  journal={Fundamental Theories of Physics},
  url={https://doi.org/10.1007/978-3-319-99046-0},
  year={2019},
  pages={Springer Cham},
  volume={~},
  publisher={Springer International Publishing},
  doi={10.1007/978-3-319-99046-0}
}

@article{Campbell2025roadmap,
  title={Roadmap on quantum thermodynamics},
  author={Campbell, Steve and D'Amico, Irene and Ciampini, Mario A and others},
  journal={arXiv preprint arXiv:2504.20145},
  year={2025},
  url={https://doi.org/10.48550/arXiv.2504.20145}
}

@article{Gherardini2024,
  title = {Quasiprobabilities in Quantum Thermodynamics and Many-Body Systems},
  author = {Gherardini, Stefano and De Chiara, Gabriele},
  journal = {PRX Quantum},
  volume = {5},
  issue = {3},
  pages = {030201},
  numpages = {38},
  year = {2024},
  month = {Sep},
  publisher = {American Physical Society},
  doi = {10.1103/PRXQuantum.5.030201},
  url = {https://link.aps.org/doi/10.1103/PRXQuantum.5.030201}
}

@article{Reeb2014,
  title={An improved Landauer principle with finite-size corrections},
  author={Reeb, David and Wolf, Michael M},
  journal={New Journal of Physics},
  volume={16},
  number={10},
  pages={103011},
  year={2014},
  publisher={IOP Publishing},
  doi= {10.1088/1367-2630/16/10/103011},
  url={https://doi.org/10.1088/1367-2630/16/10/103011}
}

@article{Santini2023,
  title = {Work statistics, quantum signatures, and enhanced work extraction in quadratic fermionic models},
  author = {Santini, Alessandro and Solfanelli, Andrea and Gherardini, Stefano and Collura, Mario},
  journal = {Phys. Rev. B},
  volume = {108},
  issue = {10},
  pages = {104308},
  numpages = {13},
  year = {2023},
  month = {Sep},
  publisher = {American Physical Society},
  doi = {10.1103/PhysRevB.108.104308},
  url = {https://link.aps.org/doi/10.1103/PhysRevB.108.104308}
}

@article{Silva2008,
  title = {Statistics of the Work Done on a Quantum Critical System by Quenching a Control Parameter},
  author = {Silva, Alessandro},
  journal = {Phys. Rev. Lett.},
  volume = {101},
  issue = {12},
  pages = {120603},
  numpages = {4},
  year = {2008},
  month = {Sep},
  publisher = {American Physical Society},
  doi = {10.1103/PhysRevLett.101.120603},
  url = {https://link.aps.org/doi/10.1103/PhysRevLett.101.120603}
}

@article{Auffeves2022,
  title = {Quantum Technologies Need a Quantum Energy Initiative},
  author = {Auff\`eves, Alexia},
  journal = {PRX Quantum},
  volume = {3},
  issue = {2},
  pages = {020101},
  numpages = {12},
  year = {2022},
  month = {Jun},
  publisher = {American Physical Society},
  doi = {10.1103/PRXQuantum.3.020101},
  url = {https://link.aps.org/doi/10.1103/PRXQuantum.3.020101}
}

@article{Campaioli2024Colloquium,
  title = {Colloquium: Quantum batteries},
  author = {Campaioli, Francesco and Gherardini, Stefano and Quach, James Q. and Polini, Marco and Andolina, Gian Marcello},
  journal = {Rev. Mod. Phys.},
  volume = {96},
  issue = {3},
  pages = {031001},
  numpages = {30},
  year = {2024},
  month = {Jul},
  publisher = {American Physical Society},
  doi = {10.1103/RevModPhys.96.031001},
  url = {https://link.aps.org/doi/10.1103/RevModPhys.96.031001}
}

@article{Hovhannisyan2013PRL,
  title = {Entanglement Generation is Not Necessary for Optimal Work Extraction},
  author = {Hovhannisyan, Karen V. and Perarnau-Llobet, Mart\'{\i} and Huber, Marcus and Ac\'{\i}n, Antonio},
  journal = {Phys. Rev. Lett.},
  volume = {111},
  issue = {24},
  pages = {240401},
  numpages = {5},
  year = {2013},
  month = {Dec},
  publisher = {American Physical Society},
  doi = {10.1103/PhysRevLett.111.240401},
  url = {https://link.aps.org/doi/10.1103/PhysRevLett.111.240401}
}

@article{giorgi2015correlation,
  title={Correlation approach to work extraction from finite quantum systems},
  author={Giorgi, Gian Luca and Campbell, Steve},
  journal={Journal of Physics B: Atomic, Molecular and Optical Physics},
  volume={48},
  number={3},
  pages={035501},
  year={2015},
  url={http://dx.doi.org/10.1088/0953-4075/48/3/035501},
  doi={10.1088/0953-4075/48/3/035501}
}

@article{Santos2019PRE,
  title = {Stable adiabatic quantum batteries},
  author = {Santos, A. C. and \c{C}akmak, B. and Campbell, S. and Zinner, N. T.},
  journal = {Phys. Rev. E},
  volume = {100},
  issue = {3},
  pages = {032107},
  numpages = {8},
  year = {2019},
  month = {Sep},
  publisher = {American Physical Society},
  doi = {10.1103/PhysRevE.100.032107},
  url = {https://link.aps.org/doi/10.1103/PhysRevE.100.032107}
}

@article{Quach2020PRApp,
  title = {Using Dark States to Charge and Stabilize Open Quantum Batteries},
  author = {Quach, James Q. and Munro, William J.},
  journal = {Phys. Rev. Appl.},
  volume = {14},
  issue = {2},
  pages = {024092},
  numpages = {9},
  year = {2020},
  month = {Aug},
  publisher = {American Physical Society},
  doi = {10.1103/PhysRevApplied.14.024092},
  url = {https://link.aps.org/doi/10.1103/PhysRevApplied.14.024092}
}

@article{Campbell2011PRA,
  title = {Propagation of nonclassical correlations across a quantum spin chain},
  author = {Campbell, S. and Apollaro, T. J. G. and Di Franco, C. and Banchi, L. and Cuccoli, A. and Vaia, R. and Plastina, F. and Paternostro, M.},
  journal = {Phys. Rev. A},
  volume = {84},
  issue = {5},
  pages = {052316},
  numpages = {8},
  year = {2011},
  month = {Nov},
  publisher = {American Physical Society},
  doi = {10.1103/PhysRevA.84.052316},
  url = {https://link.aps.org/doi/10.1103/PhysRevA.84.052316}
}

@article{LewisPRB2023,
  title = {Low-dissipation data bus via coherent quantum dynamics},
  author = {Lewis, Dylan and Moutinho, Jo\~ao P. and Costa, Antonio T. and Omar, Yasser and Bose, Sougato},
  journal = {Phys. Rev. B},
  volume = {108},
  issue = {7},
  pages = {075405},
  numpages = {9},
  year = {2023},
  month = {Aug},
  publisher = {American Physical Society},
  doi = {10.1103/PhysRevB.108.075405},
  url = {https://link.aps.org/doi/10.1103/PhysRevB.108.075405}
}

@article{Cakmak2020PRE,
  title = {Ergotropy from coherences in an open quantum system},
  author = { \c{C}akmak, B.},
  journal = {Phys. Rev. E},
  volume = {102},
  issue = {4},
  pages = {042111},
  numpages = {11},
  year = {2020},
  month = {Oct},
  publisher = {American Physical Society},
  doi = {10.1103/PhysRevE.102.042111},
  url = {https://link.aps.org/doi/10.1103/PhysRevE.102.042111}
}

@article{Strasberg2021, 
title={}, 
volume={2}, 
ISSN={2691-3399}, 
url={http://dx.doi.org/10.1103/PRXQuantum.2.030202}, 
DOI={10.1103/prxquantum.2.030202}, 
number={3}, 
volume={2},
pages={030202},
journal={PRX Quantum}, 
publisher={American Physical Society (APS)}, 
author={Strasberg, P. and Winter, A.}, year={2021}}

@article{Brand_o2015, title={}, 
volume={112}, 
ISSN={1091-6490}, 
url={http://dx.doi.org/10.1073/pnas.1411728112}, 
DOI={10.1073/pnas.1411728112}, 
number={11}, 
journal={Proceedings of the National Academy of Sciences}, 
publisher={Proceedings of the National Academy of Sciences}, 
author={Brandão, F. and Horodecki, M. and Ng, N. and Oppenheim, J. and Wehner, S.}, year={2015}, pages={3275–3279} }

@article{Bayat2011,
  title = {Initializing an unmodulated spin chain to operate as a high-quality quantum data bus},
  author = {Bayat, Abolfazl and Banchi, Leonardo and Bose, Sougato and Verrucchi, Paola},
  journal = {Phys. Rev. A},
  volume = {83},
  issue = {6},
  pages = {062328},
  numpages = {9},
  year = {2011},
  month = {Jun},
  publisher = {American Physical Society},
  doi = {10.1103/PhysRevA.83.062328},
  url = {https://link.aps.org/doi/10.1103/PhysRevA.83.062328}
}

@article{evangelakos2025rapid,
  title={Rapid charging of a two-qubit quantum battery by transverse field amplitude and phase control},
  author={Evangelakos, Vasileios and Paspalakis, Emmanuel and Stefanatos, Dionisis},
  journal={Quantum Science and Technology},
  volume={10},
  number={3},
  pages={035024},
  year={2025},
  url={https://doi.org/10.1088/2058-9565/add207},
  doi={10.1088/2058-9565/add207}
}

@article{razzoli2025cyclic,
  title={Cyclic solid-state quantum battery: thermodynamic characterization and quantum hardware simulation},
  author={Razzoli, Luca and Gemme, Giulia and Khomchenko, Ilia and Sassetti, Maura and Ouerdane, Henni and Ferraro, Dario and Benenti, Giuliano},
  journal={Quantum Science and Technology},
  volume={10},
  number={1},
  pages={015064},
  year={2025},
  publisher={IOP Publishing},
  doi={10.1088/2058-9565/ad9ed4},
  url={https://doi.org/10.1088/2058-9565/ad9ed4}
}

@article{satriani2024daemonic,
  title={Daemonic quantum battery charged by thermalization},
  author={Satriani, Matias Araya and Barra, Felipe},
  journal={Quantum Science and Technology},
  volume={9},
  number={4},
  pages={045035},
  year={2024},
  publisher={IOP Publishing},
  doi={10.1088/2058-9565/ad7316},
  url={https://doi.org/10.1088/2058-9565/ad7316}
}

@article{deMoraes2024quantum,
  title={Quantum battery supercharging via counter-diabatic dynamics},
  author={de Moraes, LFC and Duriez, Alan C and Saguia, A and Santos, Alan C and Sarandy, Marcelo S},
  journal={Quantum Science and Technology},
  volume={9},
  number={4},
  pages={045033},
  year={2024},
  publisher={IOP Publishing},
  doi={10.1088/2058-9565/ad71ed},
  url={https://doi.org/10.1088/2058-9565/ad71ed}
}

@article{Le2018Spin,
  title = {Spin-chain model of a many-body quantum battery},
  author = {Le, Thao P. and Levinsen, Jesper and Modi, Kavan and Parish, Meera M. and Pollock, Felix A.},
  journal = {Phys. Rev. A},
  volume = {97},
  issue = {2},
  pages = {022106},
  numpages = {9},
  year = {2018},
  month = {Feb},
  publisher = {American Physical Society},
  doi = {10.1103/PhysRevA.97.022106},
  url = {https://link.aps.org/doi/10.1103/PhysRevA.97.022106}
}

@article{Kamin2020PRE,
  title = {Entanglement, coherence, and charging process of quantum batteries},
  author = {Kamin, F. H. and Tabesh, F. T. and Salimi, S. and Santos, Alan C.},
  journal = {Phys. Rev. E},
  volume = {102},
  issue = {5},
  pages = {052109},
  numpages = {7},
  year = {2020},
  month = {Nov},
  publisher = {American Physical Society},
  doi = {10.1103/PhysRevE.102.052109},
  url = {https://link.aps.org/doi/10.1103/PhysRevE.102.052109}
}

@article{apollaro2015work,
  title={Work statistics, irreversible heat and correlations build-up in joining two spin chains},
  author={Apollaro, Tony JG and Francica, Gianluca and Paternostro, Mauro and Campisi, Michele},
  journal={Physica Scripta},
  volume={2015},
  number={T165},
  pages={014023},
  year={2015},
  publisher={IOP Publishing},
  url={https://doi.org/10.1088/0031-8949/2015/T165/014023},
  doi={10.1088/0031-8949/2015/T165/014023}
}

@article{moraes2022charging,
  title={Charging power and stability of always-on transitionless driven quantum batteries},
  author={Moraes, Luiz FC and Saguia, Andreia and Santos, Alan C and Sarandy, Marcelo S},
  journal={Europhysics Letters},
  volume={136},
  number={2},
  pages={23001},
  year={2022},
  publisher={IOP Publishing},
  url={https://doi.org/10.1209/0295-5075/ac1363},
  doi={10.1209/0295-5075/ac1363}
}

@article{tibben2025PRXEnergy,
  title = {Extending the Self-Discharge Time of Dicke Quantum Batteries Using Molecular Triplets},
  author = {Tibben, Daniel J. and Della Gaspera, Enrico and van Embden, Joel and Reineck, Philipp and Quach, James Q. and Campaioli, Francesco and G\'omez, Daniel E.},
  journal = {PRX Energy},
  volume = {4},
  issue = {2},
  pages = {023012},
  numpages = {10},
  year = {2025},
  month = {Jun},
  publisher = {American Physical Society},
  doi = {10.1103/bhyh-53np},
  url = {https://link.aps.org/doi/10.1103/bhyh-53np}
}

@article{Barra20219PRL,
  title = {Dissipative Charging of a Quantum Battery},
  author = {Barra, Felipe},
  journal = {Phys. Rev. Lett.},
  volume = {122},
  issue = {21},
  pages = {210601},
  numpages = {6},
  year = {2019},
  month = {May},
  publisher = {American Physical Society},
  doi = {10.1103/PhysRevLett.122.210601},
  url = {https://link.aps.org/doi/10.1103/PhysRevLett.122.210601}
}

@article{quach2022superabsorption,
  title={Superabsorption in an organic microcavity: Toward a quantum battery},
  author={Quach, James Q and McGhee, Kirsty E and Ganzer, Lucia and others},
  journal={Science advances},
  volume={8},
  number={2},
  pages={eabk3160},
  year={2022},
  publisher={American Association for the Advancement of Science},
  url={https://doi.org/10.1126/sciadv.abk3160},
  doi={10.1126/sciadv.abk3160}
}

@article{binder2015quantacell,
  title={Quantacell: powerful charging of quantum batteries},
  author={Binder, Felix C and Vinjanampathy, Sai and Modi, Kavan and Goold, John},
  journal={New Journal of Physics},
  volume={17},
  number={7},
  pages={075015},
  year={2015},
  publisher={IOP Publishing},
  url={https://doi.org/10.1088/1367-2630/17/7/075015},
  doi={10.1088/1367-2630/17/7/075015}
}

@article{Alicki2013PRE,
  title = {Entanglement boost for extractable work from ensembles of quantum batteries},
  author = {Alicki, Robert and Fannes, Mark},
  journal = {Phys. Rev. E},
  volume = {87},
  issue = {4},
  pages = {042123},
  numpages = {4},
  year = {2013},
  month = {Apr},
  publisher = {American Physical Society},
  doi = {10.1103/PhysRevE.87.042123},
  url = {https://link.aps.org/doi/10.1103/PhysRevE.87.042123}
}

@article{Ronke2011PRA,
  title = {Effect of perturbations on information transfer in spin chains},
  author = {Ronke, R. and Spiller, T. P. and D'Amico, I.},
  journal = {Phys. Rev. A},
  volume = {83},
  issue = {1},
  pages = {012325},
  numpages = {11},
  year = {2011},
  month = {Jan},
  publisher = {American Physical Society},
  doi = {10.1103/PhysRevA.83.012325},
  url = {https://link.aps.org/doi/10.1103/PhysRevA.83.012325}
}

@article{deChiara2005PRA,
  title = {From perfect to fractal transmission in spin chains},
  author = {De Chiara, Gabriele and Rossini, Davide and Montangero, Simone and Fazio, Rosario},
  journal = {Phys. Rev. A},
  volume = {72},
  issue = {1},
  pages = {012323},
  numpages = {7},
  year = {2005},
  month = {Jul},
  publisher = {American Physical Society},
  doi = {10.1103/PhysRevA.72.012323},
  url = {https://link.aps.org/doi/10.1103/PhysRevA.72.012323}
}

@article{kay2010perfect,
  title={Perfect, efficient, state transfer and its application as a constructive tool},
  author={Kay, Alastair},
  journal={International Journal of Quantum Information},
  volume={8},
  number={04},
  pages={641--676},
  year={2010},
  publisher={World Scientific},
  doi={10.1142/S0219749910006514},
  url={https://doi.org/10.1142/S0219749910006514}
}

@article{Andolina2018PRB,
  title = {Charger-mediated energy transfer in exactly solvable models for quantum batteries},
  author = {Andolina, Gian Marcello and Farina, Donato and Mari, Andrea and Pellegrini, Vittorio and Giovannetti, Vittorio and Polini, Marco},
  journal = {Phys. Rev. B},
  volume = {98},
  issue = {20},
  pages = {205423},
  numpages = {11},
  year = {2018},
  month = {Nov},
  publisher = {American Physical Society},
  doi = {10.1103/PhysRevB.98.205423},
  url = {https://link.aps.org/doi/10.1103/PhysRevB.98.205423}
}

@article{Rinaldi2025PRA,
  title = {Reliable quantum advantage in quantum battery charging},
  author = {Rinaldi, Davide and Filip, Radim and Gerace, Dario and Guarnieri, Giacomo},
  journal = {Phys. Rev. A},
  volume = {112},
  issue = {1},
  pages = {012205},
  numpages = {11},
  year = {2025},
  month = {Jul},
  publisher = {American Physical Society},
  doi = {10.1103/6kwv-z6fx},
  url = {https://link.aps.org/doi/10.1103/6kwv-z6fx}
}

@article{bezaz2025quasi,
  title={Quasi-Perfect state transfer in spin chains via parametrization of on-site energies},
  author={Bezaz, F. and Nelmes, C. C. and Estarellas, M. P. and Spiller, T. P. and D’Amico, I.},
  journal={Physica Scripta},
  volume={100},
  number={5},
  pages={055114},
  year={2025},
  publisher={IOP Publishing},
  doi = {10.1088/1402-4896/adcb71},
  url = {https://iopscience.iop.org/article/10.1088/1402-4896/adcb71}
}

@article{alsulami2024scalable,
  title={Scalable quantum spin networks from unitary construction},
  author={Alsulami, A. H. and D'Amico, I. and Estarellas, M. P. and Spiller, T. P.},
  journal={Advanced Quantum Technologies},
  volume={7},
  number={2},
  pages={2300238},
  year={2024},
  publisher={Wiley Online Library},
  doi = {10.1002/qute.202300238},
  url = {https://advanced.onlinelibrary.wiley.com/doi/10.1002/qute.202300238}
}

@book{nikolopoulos2014quantum,
  title={Quantum state transfer and network engineering},
  author={Nikolopoulos, G. M. and Jex, I. and others},
  year={2014},
  publisher={Springer},
  doi = {10.1007/978-3-642-39937-4},
  url = {https://doi.org/10.1007/978-3-642-39937-4}
}

@article{nikolopoulos2004electron,
  title={Electron wavepacket propagation in a chain of coupled quantum dots},
  author={Nikolopoulos, G. M. and Petrosyan, D. and Lambropoulos, P.},
  journal={Journal of Physics: Condensed Matter},
  volume={16},
  number={28},
  pages={4991},
  year={2004},
  publisher={IOP Publishing},
  doi = {10.1088/0953-8984/16/28/019},
  url = {https://iopscience.iop.org/article/10.1088/0953-8984/16/28/019}
}

@article{Estarellas2017PRA,
  title = {Robust quantum entanglement generation and generation-plus-storage protocols with spin chains},
  author = {Estarellas, M. P. and D'Amico, I. and Spiller, T. P.},
  journal = {Phys. Rev. A},
  volume = {95},
  issue = {4},
  pages = {042335},
  numpages = {7},
  year = {2017},
  month = {Apr},
  publisher = {American Physical Society},
  doi = {10.1103/PhysRevA.95.042335},
  url = {https://link.aps.org/doi/10.1103/PhysRevA.95.042335}
}

@article{DiFranco2008PRL,
  title = {Perfect State Transfer on a Spin Chain without State Initialization},
  author = {Di Franco, C. and Paternostro, M. and Kim, M. S.},
  journal = {Phys. Rev. Lett.},
  volume = {101},
  issue = {23},
  pages = {230502},
  numpages = {4},
  year = {2008},
  month = {Dec},
  publisher = {American Physical Society},
  doi = {10.1103/PhysRevLett.101.230502},
  url = {https://link.aps.org/doi/10.1103/PhysRevLett.101.230502}
}

@article{Apollaro2012PRA,
  title = {99$%$-fidelity ballistic quantum-state transfer through long uniform channels},
  author = {Apollaro, T. J. G. and Banchi, L. and Cuccoli, A. and Vaia, R. and Verrucchi, P.},
  journal = {Phys. Rev. A},
  volume = {85},
  issue = {5},
  pages = {052319},
  numpages = {11},
  year = {2012},
  month = {May},
  publisher = {American Physical Society},
  doi = {10.1103/PhysRevA.85.052319},
  url = {https://link.aps.org/doi/10.1103/PhysRevA.85.052319}
}

@article{banchi2011long,
  title={Long quantum channels for high-quality entanglement transfer},
  author={Banchi, L. and Apollaro, T. J. G. and Cuccoli, A. and Vaia, R. and Verrucchi, P.},
  journal={New Journal of Physics},
  volume={13},
  number={12},
  pages={123006},
  year={2011},
  publisher={IOP Publishing},
  doi = {10.1088/1367-2630/13/12/123006},
  url = {https://iopscience.iop.org/article/10.1088/1367-2630/13/12/123006}
}

@article{Vinet2012PRA,
  title = {How to construct spin chains with perfect state transfer},
  author = {Vinet, Luc and Zhedanov, Alexei},
  journal = {Phys. Rev. A},
  volume = {85},
  issue = {1},
  pages = {012323},
  numpages = {7},
  year = {2012},
  month = {Jan},
  publisher = {American Physical Society},
  doi = {10.1103/PhysRevA.85.012323},
  url = {https://link.aps.org/doi/10.1103/PhysRevA.85.012323}
}

@article{wilkinson2017rapid,
  title={Rapid and Robust generation of Einstein--Podolsky--Rosen pairs with Spin Chains},
  author={Wilkinson, Kieran N and Estarellas, Marta P and Spiller, Timothy P and D'amico, Irene},
  journal={Quant. Inf. Comput.},
  volume={18},
  pages = {0247-0264},
  year={2017},
  url={https://doi.org/10.48550/arXiv.1708.05650},
  doi={https://doi.org/10.26421/QIC18.3-4-5}
}

@article{Christandl2017,
   title={},
   volume={96},
   ISSN={2469-9934},
   url={http://dx.doi.org/10.1103/PhysRevA.96.032335},
   DOI={10.1103/physreva.96.032335},
   number={3},
   journal={Phys. Rev. A},
   publisher={American Physical Society (APS)},
   author={Christandl, M. and Vinet, L. and Zhedanov, A.},
   year={2017} }

@misc{mondal2025,
      title={}, 
      author={Shuva, M. and Debarupa, S. and Ujjwal, S.},
      year={2025},
      eprint={2507.16610},
      archivePrefix={arXiv},
      primaryClass={quant-ph},
      url={https://arxiv.org/abs/2507.16610}, 
}

@article{Tirone2023,
  title = {Work Extraction Processes from Noisy Quantum Batteries: The Role of Nonlocal Resources},
  author = {Tirone, S. and Salvia, R. and Chessa, S. and Giovannetti, V.},
  journal = {Phys. Rev. Lett.},
  volume = {131},
  issue = {6},
  pages = {060402},
  numpages = {6},
  year = {2023},
  publisher = {American Physical Society},
  doi = {10.1103/PhysRevLett.131.060402},
  url = {https://link.aps.org/doi/10.1103/PhysRevLett.131.060402}
}

@article{Tirone2021,
  title = {Quantum Energy Lines and the Optimal Output Ergotropy Problem},
  author = {Tirone, S. and Salvia, R. and Giovannetti, V.},
  journal = {Phys. Rev. Lett.},
  volume = {127},
  issue = {21},
  pages = {210601},
  numpages = {7},
  year = {2021},
  publisher = {American Physical Society},
  doi = {10.1103/PhysRevLett.127.210601},
  url = {https://link.aps.org/doi/10.1103/PhysRevLett.127.210601}
}

\end{document}